\documentclass[prd,twocolumn,amsmath,amssymb,floatfix,superscriptaddress]{revtex4}
\usepackage{bm}
\usepackage{amsmath}
\usepackage{epsf}
\usepackage{color}
\usepackage{natbib}
\usepackage{graphicx}
\def\simlt{\lesssim}
\def\simgt{\gtrsim}

\newcommand{\rhom}{\rho_{\rm m}}
\newcommand{\Mpch}{\mbox{Mpc}/h}
\newcommand{\iMpch}{h/\mbox{Mpc}}
\newcommand{\Msun}{M_\odot}

\newcommand{\refeq}[1]{Eq.~(\ref{eq:#1})}
\newcommand{\refssec}[1]{section~\ref{subsec:#1}}
\newcommand{\reffig}[1]{Fig.~\ref{plot:#1}}
\newcommand{\refFig}[1]{Fig.~\ref{plot:#1}}

\definecolor{darkgreen}{cmyk}{0.85,0.2,1.00,0.2} 
\definecolor{purple}{cmyk}{0.5,1.0,0,0}

\begin{document}
\title{Non-linear Evolution of $\bm{f(R)}$ Cosmologies III: Halo Statistics}

\author{Fabian Schmidt}
\email{fabians@uchicago.edu}
\affiliation{Department of Astronomy \& Astrophysics, University of Chicago, Chicago IL 60637}
\affiliation{Kavli Institute for Cosmological Physics, University of Chicago, Chicago IL 60637}
\author{Marcos Lima}
\affiliation{Kavli Institute for Cosmological Physics, University of Chicago, Chicago IL 60637}
\affiliation{Department of Physics, University of Chicago, Chicago IL 60637}
\affiliation{Department of Physics \& Astronomy, University of Pennsylvania, Philadelphia PA 19104}

\author{Hiroaki Oyaizu}
\affiliation{Department of Astronomy \& Astrophysics, University of Chicago, Chicago IL 60637}
\affiliation{Kavli Institute for Cosmological Physics, University of Chicago, Chicago IL 60637}

\author{Wayne Hu}
\affiliation{Department of Astronomy \& Astrophysics, University of Chicago, Chicago IL 60637}
\affiliation{Kavli Institute for Cosmological Physics, University of Chicago, Chicago IL 60637}
\affiliation{Enrico Fermi Institute, University of Chicago, Chicago IL 60637}

\date{\today}

\begin{abstract}
\baselineskip 11pt
The statistical properties of dark matter halos, the building blocks 
of cosmological observables associated
with structure in the universe, offer many opportunities to test models for 
cosmic acceleration,
especially those that seek to modify gravitational forces.   We study the
abundance, bias and profiles of halos in cosmological simulations for one such model: the modified action $f(R)$ theory.
The effects of $f(R)$ modified gravity can be separated into a large- and
small-field limit.
In the large field limit, which is accessible to current observations, 
enhanced gravitational
forces raise the abundance of rare massive halos and decrease their bias but leave their
(lensing) mass profiles largely unchanged.  This regime is well described by scaling relations based on a modification of spherical collapse
calculations.   In the small field limit, the enhancement of the gravitational
force is suppressed inside
halos and the effects on halo properties are substantially 
reduced for the most massive halos.  Nonetheless,
the scaling relations still retain limited applicability for the purpose of establishing
conservative upper limits on the modification to gravity.
\end{abstract}

\maketitle

\section{Introduction} \label{sec:intro}
In the so-called $f(R)$ class of models (see \cite{Sotiriou:2008rp,Nojiri:2008nt} and references therein) cosmic acceleration arises not from an exotic form of energy with negative pressure
but from a modification of gravity that replaces the Einstein-Hilbert action by a function of
the Ricci or curvature scalar $R$  \cite{Caretal03,NojOdi03,Capozziello:2003tk}.

Cosmological simulations are crucial for exposing the phenomenology of $f(R)$ models. 
In order to satisfy local tests of gravity, $f(R)$ models exhibit a non-linear process, called the
chameleon mechanism, 
to suppress force modifications in the deep potential wells of cosmological structure
 \cite{khoury04a,Cembranos:2005fi,Navarro:2006mw,Faulkner:2006ub,HuSaw07a}.   
Upcoming tests of cosmic acceleration from gravitational lensing, galaxy and cluster
surveys have most of their
statistical weight in the weakly to fully non-linear regime. Stringent constraints
on modified gravity can be expected from current and future surveys, once
the impact on observables in the non-linear regime is understood.

In the previous papers  in this series, we have established the methodology for
cosmological $f(R)$ simulations \cite{oyaizu08b} and conducted a suite 
of simulations that uncover the chameleon mechanism and its effect on the
matter power spectrum \cite{Pkpaper}. 
In this paper, we continue our exploration of the non-linear aspects of the
$f(R)$ model
by examining the properties of the basic building blocks of cosmological structure: dark matter halos.   Specifically, we quantify their abundance, i.e. the
halo mass function, clustering properties, i.e. the linear bias, and their
density profiles,
to see how each are modified from the standard cosmological constant, cold dark
matter model $\Lambda$CDM.

We begin in \S \ref{sec:meth} with a brief review of the important properties of $f(R)$ models and a discussion of the simulation and analysis methodology.  
 We present our results on halo statistics in \S \ref{sec:results} and discuss them in \S \ref{sec:discussion}. 
 Throughout we place a special emphasis on exploring the impact of the
chameleon mechanism and highlighting differences between the simulations and
conventional scaling relations based on linear theory and  $\Lambda$CDM.
These differences expose crucial distinctions that must be considered when
observationally testing modified gravity theories.

\section{Methods} \label{sec:meth}

We begin in \S \ref{subsec:fr} 
by briefly reviewing the basic properties of the $f(R)$ model that are important
for understanding the cosmological simulations described in \S \ref{subsec:sim}.   
We refer
the reader to \cite{Pkpaper} for a more detailed treatment.  
Finally in \S \ref{subsec:halo}, we discuss the methods used in identifying the halos
and measuring their abundance, bias and profiles. 

\subsection{$\bm{f(R)}$ Gravity} \label{subsec:fr}

The $f(R)$ model generalizes the
Einstein-Hilbert action to include an arbitrary function of the scalar curvature $R$,
 \begin{eqnarray}
S =  \int{d^4 x \sqrt{-g} \left[ \frac{R+f(R)}{16\pi G} + L_{m} \right]}\,. 
\label{eqn:action}
\end{eqnarray}
Here $L_{m}$ is the Lagrangian of the ordinary 
matter and throughout $c=\hbar=1$.    Force modifications are associated
with an additional scalar degree of freedom $f_{R}\equiv df/dR$.
For definiteness, we choose the functional form for $f(R)$ given in \cite{HuSaw07a} (with $n=1$), but neglect higher corrections of order $|f_{R0}| \leq 10^{-4}$ which 
results in the following effective $f(R)$:
\begin{eqnarray}
f(R) = -16 \pi G \rho_\Lambda - f_{R0} \frac{\bar R_0^2}{ R} \,.
\label{eqn:fRapprox}
\end{eqnarray}
  Here we define $\bar R_{0}=\bar R(z=0)$ 
and $f_{R0}= f_{R}(\bar R_{0})$, where overbars denote the quantities of the
background spacetime.  
For $|f_{R0}| \ll 1$ the background expansion history mimics $\Lambda$CDM with
$\Omega_\Lambda = \rho_\Lambda/\rho_{\rm crit}$. 

Variation of Eq.~(\ref{eqn:action}) with respect to the metric yields the 
modified Einstein equations. 
We work in the quasistatic limit, where time derivatives may be neglected 
compared with spatial derivatives. 
The trace of the modified Einstein equations yields
the $f_R$ field equation
\begin{eqnarray}
\nabla^2 \delta  f_{R} = \frac{a^{2}}{3}\left[\delta R(f_{R}) - 8 \pi G \delta \rho_{\rm m}\right] \,,\label{eqn:frorig}
\end{eqnarray}
where coordinates are comoving, $\delta f_R = f_R(R)-f_R(\bar R)$,
$\delta R = R -\bar R$, $\delta \rho_{\rm m} = \rho_{\rm m} - \bar \rho_{\rm m}$.
The time-time component of the Einstein equations yields the modified Poisson 
equation
\begin{eqnarray}
\nabla^2 \Psi = \frac{16 \pi G}{3}a^{2} \delta \rho_{\rm m} - \frac{a^{2}}{6} \delta R(f_R) \,.\label{eqn:potorig}
\end{eqnarray}
Here $\Psi$ is the Newtonian potential or time-time metric perturbation $2\Psi = \delta g_{00}/g_{00}$ in the longitudinal gauge.    These two equations define a closed system for
the Newtonian potential given the density field.    The matter falls in the Newtonian potential
as usual and so the modifications to gravity are completely contained in the equation for $\Psi$.

The field equation~(\ref{eqn:frorig}) is a non-linear
Poisson-type equation, where the non-linearity is determined by $\delta R(f_R)$.
If the background field $f_{R0}$ is sufficiently large, then field fluctuations are
relatively small and this term may be linearized as  $\delta R \approx (d R/df_R)|_{\bar R}\: \delta f_R$.
It is straightforward to show that the Fourier space solution to Eqs.~(\ref{eqn:frorig}) 
and (\ref{eqn:potorig}) in this approximation is
\begin{equation}
k^2 \Psi({\bf k})  = - 4\pi G \left( \frac{4}{3} - \frac{1}{3} \frac{\mu^2 a^2}{k^2 + \mu^2 a^2}\right)
a^2 \delta \rho_{\rm m}({\bf k}) \, , \label{eqn:linearfr}
\end{equation}
with $\mu = (3 d f_R/dR)^{-1/2}$.  Hence, gravitational forces are enhanced
by a factor of $4/3$ on scales below $\mu^{-1}$, the Compton wavelength of the field.
We call this regime the \textit{large field limit}.

Eqs.~(\ref{eqn:frorig}) and (\ref{eqn:potorig}) in the large field limit imply that the
field fluctuations  are of order the gravitational potential $|\delta f_R| \sim |\Psi|$.  
Therefore if the background field is of order the typical gravitational potentials
of cosmological structure $|\Psi| \simlt 10^{-5}$ or smaller, field fluctuations
become of order unity and 
$\delta R \gg  (d R/df_R) \delta f_R$ which causes the Compton wavelength to 
shrink \cite{HuSaw07a}.
We call this the \textit{small field limit}.   The large and small field limits 
are separated by a value of the background field of $|f_{R0}| \sim 10^{-5}$.

In the small field limit, the field equation~(\ref{eqn:frorig}) then requires $\delta R \approx 8\pi G 
\delta \rho_{\rm m}$ which drives the Poisson equation~(\ref{eqn:potorig}) back to its usual form.
This is the so-called chameleon mechanism which occurs when the background field
is small compared with the depth of the gravitational potential.    Hence force law
deviations are suppressed in the deepest gravitational potentials, i.e.\ inside
the high  overdensities of collapsed dark matter halos.

It is important to note that due to the modified Poisson equation~(\ref{eqn:potorig}) for the dynamical potential, the masses dealt with in this paper 
correspond observationally to gravitational lensing masses,
and {\it not} to dynamical masses (see Appendix \ref{app:sph_collapse}).

\subsection{Simulations} \label{subsec:sim}

To solve the system of equations defined by the modified Poisson equation
(\ref{eqn:potorig}) and the $f_R$ field equation (\ref{eqn:frorig}) in the context
of cosmological structure formation, 
we employ the methodology described in \cite{oyaizu08b} and implemented in
 \cite{Pkpaper}.
Briefly, the field equation for
 $f_R$ is solved on a regular
grid using relaxation techniques and multigrid iteration \citep{brandt73,briggs00a}.
The potential $\Psi$ is computed from the density and $f_R$ fields
using the fast Fourier transform method.
The dark matter particles are then moved according to the gradient of the
computed potential, $-\nabla \Psi$, using a second order accurate leap-frog integrator.

We choose a range of background field values $|f_{R0}|= 10^{-6}-10^{-4}$ to expose the
impact of the chameleon mechanism.   Since cosmological potentials range from $10^{-6}-
10^{-5}$, we expect the chameleon mechanism to be operative in the small field limit
of this range but absent in the large field limit.   We also include $|f_{R0}|=0$ which is
equivalent to $\Lambda$CDM.  Note that the background expansion history for all
runs are indistinguishable from $\Lambda$CDM to ${\cal O}(f_{R0})$.
More specifically, we take a flat background cosmology defined by $\Omega_\Lambda=0.76$,
$\Omega_b=0.04181$, $H_0=73$ km/s/Mpc and initial power in curvature fluctuations
$A_s=(4.73\times 10^{-5})^2$ at $k=0.05$Mpc$^{-1}$ with a tilt of $n_s=0.958$.

To more directly assess the impact of the chameleon mechanism, we also carry out linearized $f_R$
simulations in which the gravitational potential, $\Psi$, is evaluated according to
Eq.~(\ref{eqn:linearfr}).
In the linearized treatment, the Compton wavelength is assumed to be
fixed by the background field and thus chameleon effects are not present.  Therefore, the difference between the full $f_R$ simulations and the linearized $f_R$
simulations are wholly due to the chameleon effects. We will call these runs the 
``no-chameleon" simulations.  

 Table~\ref{table:runs} lists the properties of the simulations
used in the analysis below.   All simulations possess 512 grid cells in each direction and 
$N_p=256^3$ particles.

\begin{table}
\caption{Simulation type and number of runs per box size. } 
\begin{center}
  \leavevmode
  \begin{tabular}{c|c|c c c c}
  && \multicolumn{3}{|c}{$L_{\rm box}$ ($h^{-1}$ Mpc)} \\ 
  \cline{3-6} 
  
&$|f_{R0}|$ &\ \ $400$\ \ & $256 $\ \ \  & $128$\ \ \  & $64 $\ \ \  \\
\hline
\# of\ \ & $10^{-4}$ & 6 & 6 & 6 & 6\\
boxes\ \ & $10^{-5}$ & 6 & 6 & 6 & 6\\
     & $10^{-6}$ & 6 & 6 & 6 & 6\\
&0 ($\Lambda$CDM) & 6     & 6     & 6     & 6 \\
\hline
\multicolumn{2}{c|}{$M_{\rm h, min}$ ($10^{12} h^{-1} M_\odot$)\ \ } & 204  & 53.7 & 6.61 & 0.83 \\
\hline
\multicolumn{2}{c|}{$k_{\rm fun}=\pi/L_{\rm box}$  ($h$ Mpc$^{-1}$)\ \            } & 0.008 & 0.012 & 0.025 & 0.049 \\
\hline
\multicolumn{2}{c|}{$r_{\rm cell}$ ($h^{-1}$ Mpc)\ \            } & 0.78 & 0.50 & 0.25 & 0.125 \\
\hline
\end{tabular}
\end{center}
\label{table:runs}
\end{table}

\subsection{Halo Properties} \label{subsec:halo}

We  identify halos
and measure their masses in simulations with a spherical overdensity algorithm similar to \cite{Jenkins01}.
We use cloud-in-cell interpolation to assign the particles
to the grid.  Starting at the highest overdensity grid point, we then count the
particles within a growing sphere centered on the
center of mass, until the desired overdensity with respect to the mean matter density 
$\Delta_{\rm th}=\rho_{\rm m}/\bar\rho_{\rm m}$ is reached. Here, we take
$\Delta_{\rm th}=300$ for definiteness.  The mass $M_{300}$ of
the halo is then defined by the mass of all particles enclosed within this 
radius $r_{300}$.
We move onto the next highest density grid cell and repeat the procedure until 
all halos have been identified.
We implicitly take $M=M_{300}$ below unless otherwise 
specified.

In our final results we only keep halos with
at least $N_{\rm min}$ dark matter particles, and
since our simulations are not of high-resolution, 
we conservatively take $N_{\rm min}=800$. We verified that a lower minimum
particle number of $N_{\rm min}=400$
provides results consistent with statistical uncertainties for all our 
quoted halo properties.
The corresponding minimum 
masses of halos are listed in Table~\ref{table:runs}.

For each simulation run, we determine the halo mass function by binning
halos in logarithmic mass intervals, and dividing by the comoving
volume of the simulation box. We then combine different runs and box sizes
using a bootstrap procedure
to produce the estimate of the mass function and its errors.
We weight each box by volume and use
only those boxes whose minimum halo mass is below the mass bin considered.
 When measuring differences 
between $\Lambda$CDM and $f(R)$, we average the differences 
between simulations with the same initial conditions to reduce the sample 
variance. 

We compare simulation results to 
the Sheth-Tormen (ST) prescription \cite{SheTor99} given in Appendix \ref{subsec:scaling}
 with modifications to spherical
collapse as detailed in the Appendix \ref{app:sph_collapse}.  Semi-analytic prescriptions of this
type are widely used when analyzing data for cosmological constraints and
so an assessment of their range of validity is of practical importance.

Next we extract the {\it linear} halo bias
$b_{\rm L}(M)$ from our simulations. 
For halos of a given 
logarithmic mass range in a box of size $L_{\rm box}$, we first obtain the 
halo bias $b(k,M)$ by dividing the halo-mass cross spectrum by the matter 
power spectrum for each simulation
\begin{eqnarray}
b(k,M) = \frac{P_{\rm hm}(k,M)}{P_{\rm mm}(k)}
       =\frac{\langle \delta_{\rm h}^*({\bf k},M) \delta_{}({\bf k}) \rangle_{\bf k}}{\langle \delta_{}^*(
       {\bf k}) \delta_{}({\bf k})
       \rangle_{\bf k}}\,,
\label{eqn:defbias}
\end{eqnarray}   
where $\delta_{\rm h}({\bf k},M)$ is the halo {\it number} density contrast whereas 
$\delta_{}({\bf k})$ is the matter {\it mass} density contrast.   The average is over
the $k$-modes in a $k$-bin. 
For each box we employ the modes $k \ge k_{\rm min} = 2 k_{\rm fun}$, where $k_{\rm fun}$
is the fundamental mode of the box (see Table~\ref{table:runs})  and thus
the smallest boxes barely probe the linear regime. For the larger mass bins, we 
probe more of the linear regime but are more limited by small statistical samples.
  Note that the definition of bias adopted will differ from
alternate choices such as $(P_{\rm hh}/P_{\rm mm})^{1/2}$ or $P_{\rm hh}/P_{\rm hm}$
in the non-linear regime where the correlation coefficient between halos
and matter can differ from unity.

In order to remove trends from the non-linearity of the bias, we fit a linear relation to 
$b(k,M)=a_0(M)+a_1(M) k$ between $k_{\rm min}$ and $10 k_{\rm min}$, where 
$b(k,M)$ is the combined measurement from all boxes. The linear halo 
bias in this mass range is then extrapolated as $b_{\rm L}(M)=b(k=0,M)=a_0(M)$. 
When considering the modifications in the $f(R)$ simulations, the same bootstrap and linear fit procedure is applied but to the quantity
$\Delta b/b \equiv (b_{f(R)}-b_{\Lambda\rm CDM})/b_{\Lambda\rm CDM}$.   Again we compare these results to the peak-background split predictions
based on the ST mass function detailed in Appendix \ref{subsec:scaling}.

Finally, we stack the halos in each mass interval and measure the average 
density profile and mass correlations of the halos.    To reduce scatter
within the mass bin we scale each density profile to its own $r_{300}$ before 
stacking, i.e. we measure 
\begin{equation}
\delta_\rho(r/r_{300})  \equiv \left\langle\frac{\rho_{\rm h}(r/r_{300})}{\bar\rho_{\rm m}} - 1 \right\rangle_{\rm h} \,.
\label{eqn:haloprofile}
\end{equation}
The spatial resolution of our particle-mesh simulations is limited by the
fixed size of grid cells $r_{\rm cell}$ (see Tab.~\ref{table:runs}). 
We measure halo profiles down to the grid scale, though we
expect that profiles have converged only at scales of several grid cells.  
When the resolution becomes too low, the inner profile flattens leading to a 
misestimation of both the mass enclosed at $r_{300}$ and the shape of the 
halo profiles. We therefore use only the highest resolution boxes for our 
comparisons with the $f(R)$ simulations. The maximum radius for each profile
is set to $0.4\:L_{\rm box}$.

In order to avoid biases from incompleteness effects, we further limit
the range of the stacked profile to radii where more than 90\% of the halos
in the mass bin contribute. We then bootstrap over all halos in the given
mass range in order to determine the average profile and its error.
Profiles and the halo-mass correlation function results are compared
to the Navarro-Frenk-White (NFW) profile and halo model respectively
(see Appendix \ref{subsec:scaling}).

\begin{figure}[ht!]
   \begin{center}
     \centerline{\epsfxsize=3.4in\epsffile{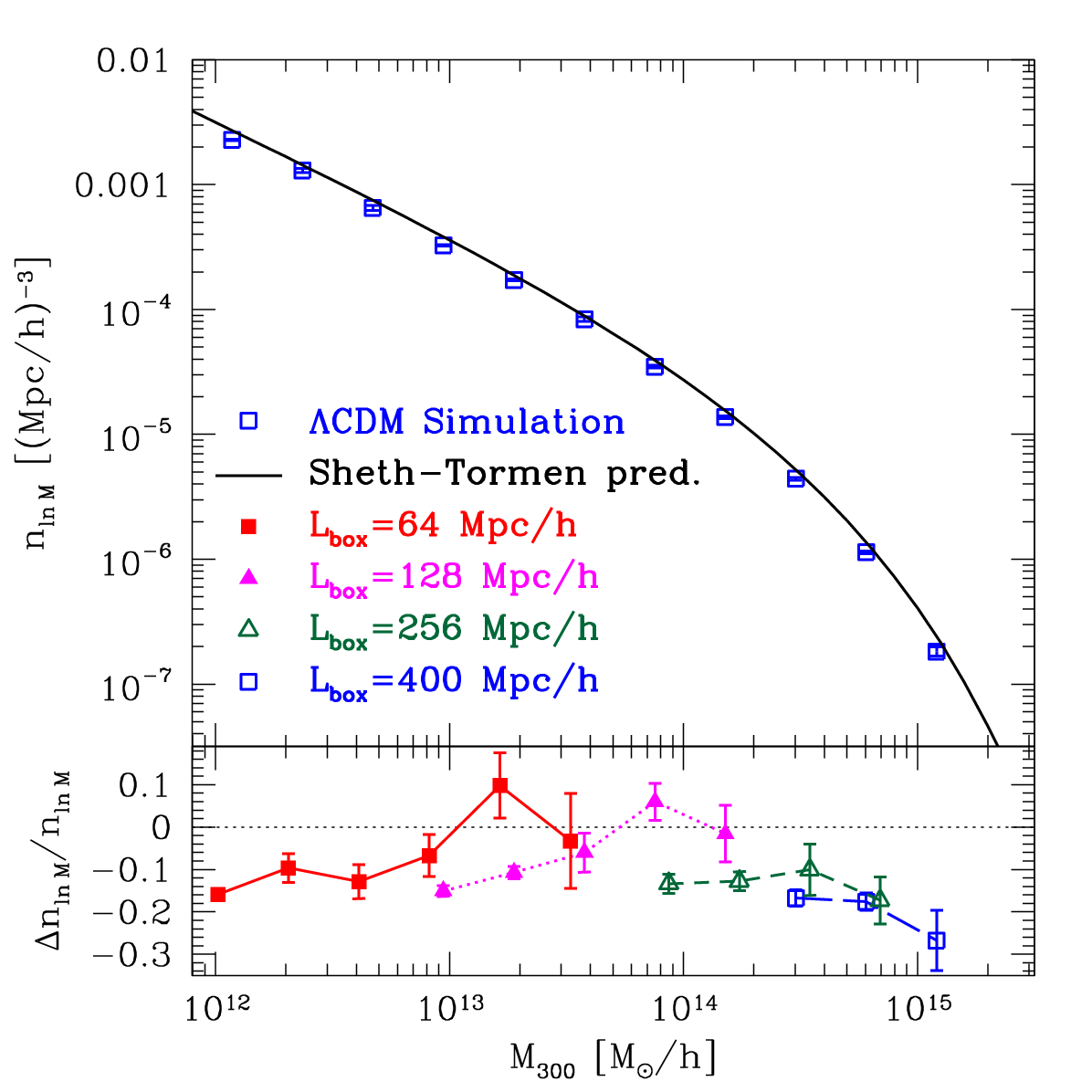}}
   \end{center}
  \caption{\footnotesize
The halo mass function as a function of $M_{300}$ measured in $\Lambda$CDM 
simulations with bootstrap errors on the mean.   The upper panel combines different box sizes from $64$ to $400\Mpch$ and 
compares results with the Sheth-Tormen prediction rescaled from $M_{\rm v}$ to $M_{300}$ as described in the text.
The lower panel shows the relative deviations from this prediction
separately for different box sizes.} 
  \label{plot:mfGR}
\end{figure}

\begin{figure}[htb!]
   \begin{center}
     \centerline{\epsfxsize=3.4in\epsffile{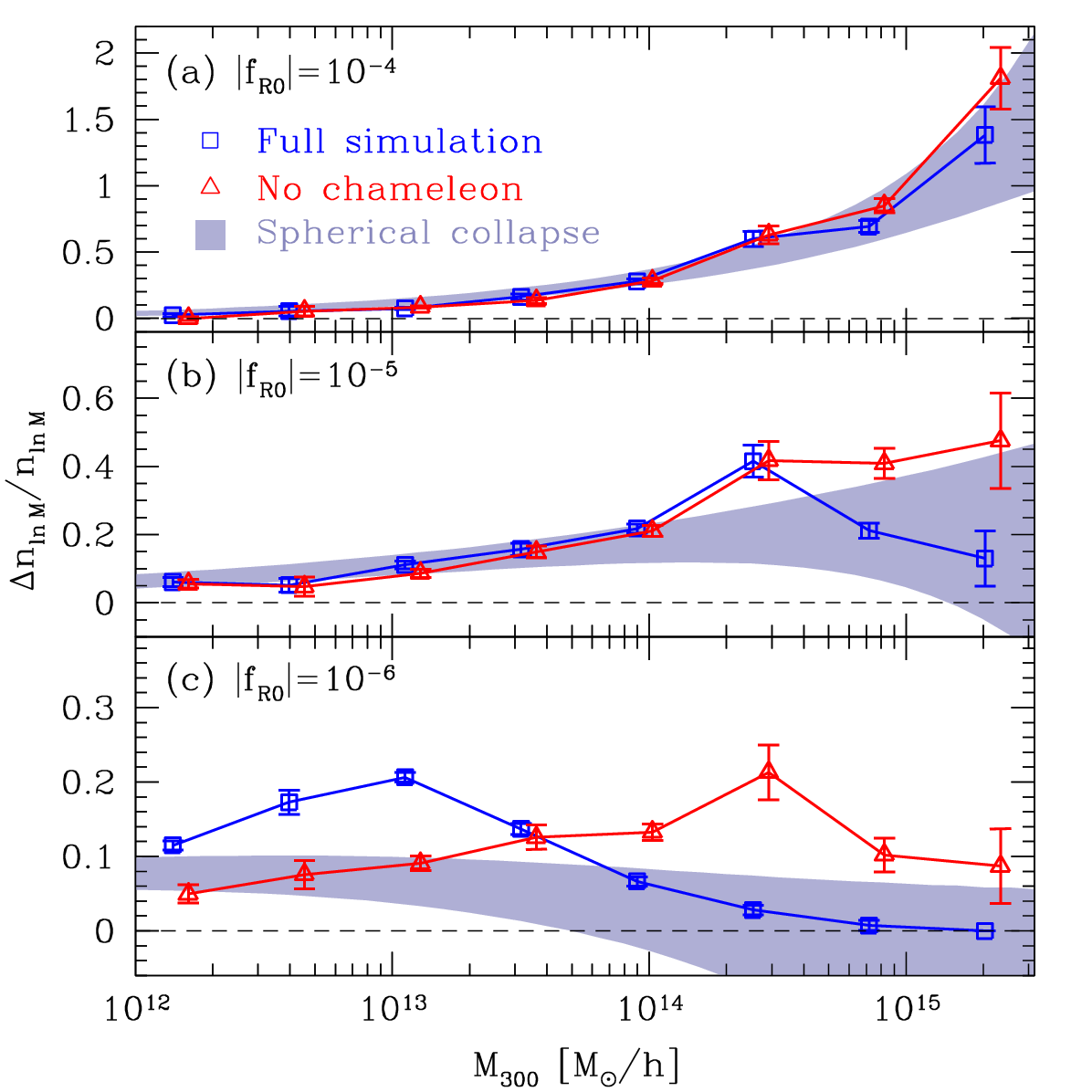}}
   \end{center}
  \caption{ \footnotesize
Relative deviations of the $f(R)$ halo mass functions from $\Lambda$CDM, with
$|f_{R0}|=10^{-4}$ (top panel), $10^{-5}$ (middle panel), and $10^{-6}$ (lower panel).
In each case, blue squares denote the full simulations, while red triangles
(displaced horizontally for visibility) denote the no chameleon simulations.
 The shaded band
shows the range of enhancement expected from spherical collapse
rescaled from $M_{\rm v}$ to $M_{300}$. 
} 
  \label{plot:diff.mf}
\end{figure}

\section{Results} \label{sec:results}

In this section we present the results obtained from N-body
simulations of the $f(R)$ models for the halo mass function
(\S \ref{subsec:dndm}), halo bias (\S \ref{subsec:bias}),
density profiles (\S \ref{subsec:prof}) and matter power spectrum
(\S \ref{subsec:power}). 
In all cases, we compare the simulation
results with predictions using scaling relations based on spherical collapse 
calculations, the Press-Schechter prescription and findings from simulations 
of $\Lambda$CDM. These calculations are detailed in the Appendices.

Since  spherical collapse predictions depend on the
gravitational force modification, we give a range of predictions in each case.
The extremes are given by collapse with standard gravity and with
enhanced forces throughout. The former follows the $\Lambda$CDM
expectation of a linear density extrapolated to collapse of
$\delta_{c}=1.673$ and a virial overdensity of $\Delta_{\rm v}=390$; the
latter modifies these parameters to $\delta_c=1.692$
and $\Delta_{\rm v}=309$ as detailed in Appendix~\ref{app:sph_collapse}.

Neither assumption for the nonlinear collapse
is completely valid given the evolving Compton wavelength and the chameleon mechanism. 
Moreover, the evolution of linear density perturbations used as the reference for
the scaling relations in Eqs.~(\ref{eqn:massfn}), (\ref{eqn:bias}), (\ref{eq:Pkhalo}), and
(\ref{eq:Phm}) assumes in both cases 
the full linear growth of the $f(R)$ model through $\sigma(M)$,
including the effects of the evolving background Compton wavelength but not the
chameleon mechanism.   Thus unmodified
spherical collapse parameters do not equate to unmodified spherical collapse
predictions.

\begin{figure}[tb]
   \begin{center}
     \centerline{\epsfxsize=3.4in\epsffile{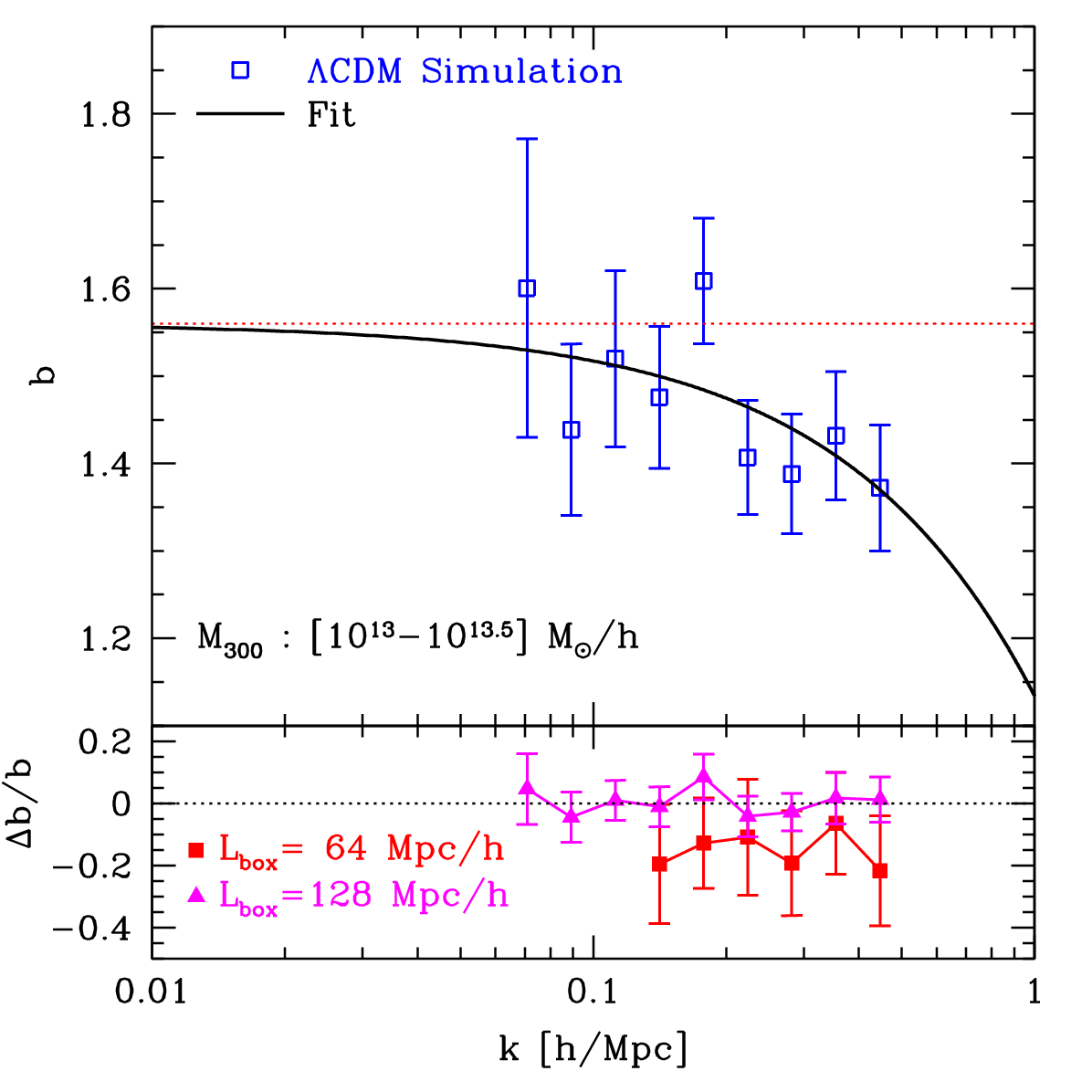}}
   \end{center}
  \caption{\footnotesize
The halo bias as a function of wavenumber $k$ in $\Lambda$CDM.  The upper panel
combines 
different box sizes and runs for halos with mass $M_{300}=10^{13}-10^{13.5} h^{-1}M_{\odot}$. 
The black solid line indicates a linear fit, whose extrapolation to $k=0$
gives $b_{\rm L}$ (dotted red line). Error bars denote bootstrap errors on the mean.
The lower panel shows the relative deviations from the fit separately for 
each box contributing in this mass range.
} 
  \label{plot:bkGR}
\end{figure}
\begin{figure}[tb]
   \begin{center}
     \centerline{\epsfxsize=3.4in\epsffile{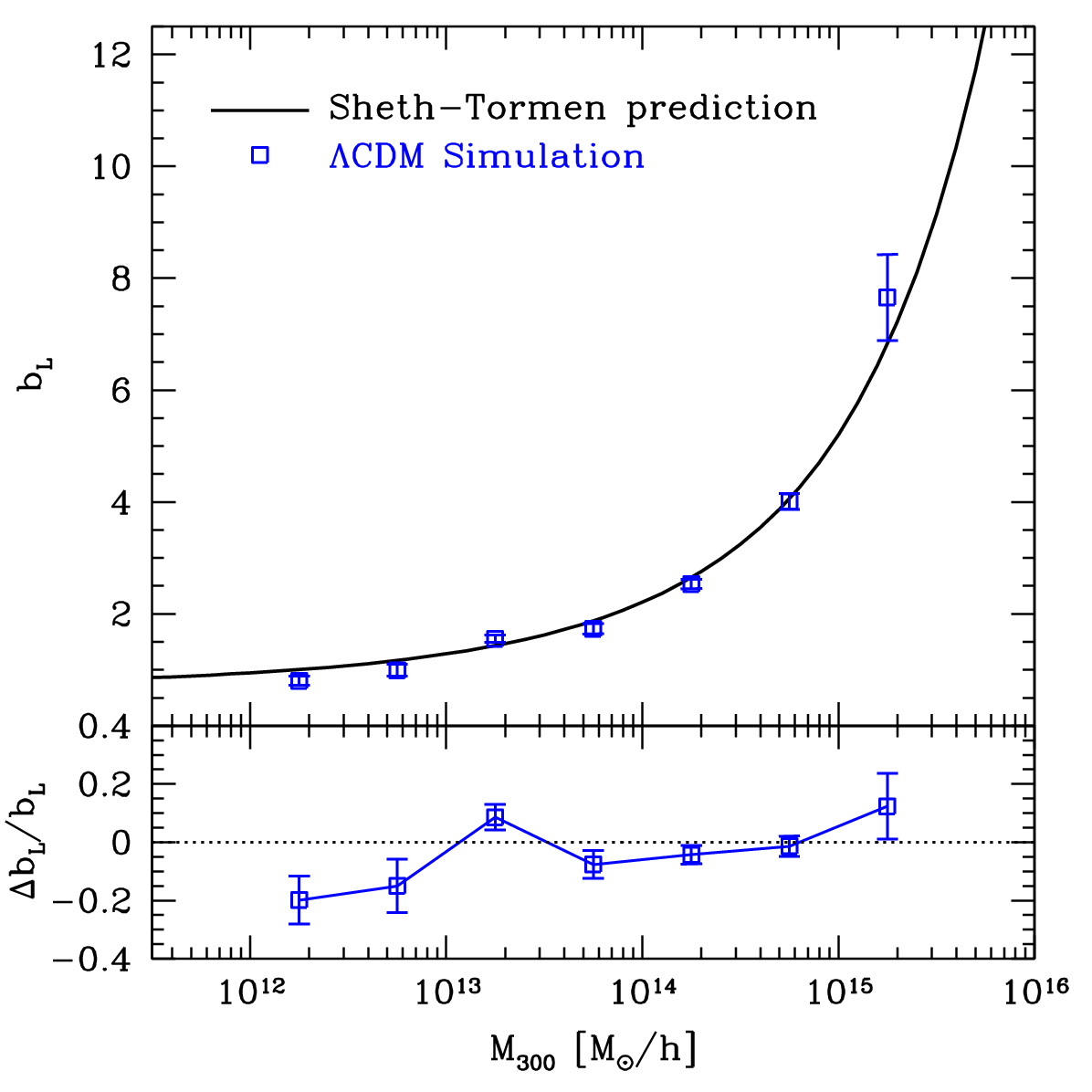}}
   \end{center}
  \caption{\footnotesize
The {\it linear} halo bias as a function of  $M_{300}$ extrapolated from the $\Lambda$CDM 
simulations with bootstrap errors on the mean.  The upper panel combines different box sizes and runs and compares the
result to the Sheth-Tormen prediction rescaling masses from $M_{\rm v}$ to $M_{300}$.
The lower panel shows the relative deviations from this prediction.
} 
  \label{plot:b1GR}
\end{figure}

\subsection{Mass Function} \label{subsec:dndm}

In Fig.~\ref{plot:mfGR}, we show the halo mass function measured from our suite of $\Lambda$CDM 
simulations along with the  bootstrap errors
described in \S \ref{subsec:halo}.  For reference, we compare the simulations to
the Sheth-Tormen (ST) mass function of Eq.~(\ref{eqn:massfn}). 
The ST formula gives the mass function in terms of the virial mass and we 
rescale it to $M_{\rm 300}$ assuming an NFW profile 
(see Appendix \ref{subsec:scaling}).
Our $\Lambda$CDM simulations are consistent with the 10-20\% level of accuracy 
expected of 
the ST formula and internally between boxes of differing resolution. 

Next, we compare the $f(R)$ and $\Lambda$CDM simulations.
Our measurement of  the halo mass function itself
is limited by statistics and to a lesser extent, resolution (see Fig.~\ref{plot:mfGR}).
 However, we can reduce the impact of both effects
by considering the relative difference between the halo
mass functions measured in $f(R)$ and $\Lambda$CDM simulations with
the same initial conditions and resolution.

Fig.~\ref{plot:diff.mf} shows this relative enhancement of the halo 
mass-function in the $f(R)$ simulations for different values of $f_{R0}$, the 
background  field today, combining different box sizes as described in \refssec{halo}. 
We show results for the full simulations as well as the no-chameleon
simulations to help highlight the impact of the chameleon mechanism. 

For the large field value of $|f_{R0}|=10^{-4}$, the number of halos
increases significantly, especially at the high mass end, 
by up to 50$-$150\% for cluster-sized halos.  The chameleon
effect slightly suppresses the abundance in the high mass end.  A similar effect
occurs for the power spectrum \cite{Pkpaper} and arises due to the appearance
of the chameleon effect in deep potentials at high redshifts where the
background field values are smaller.
The overall trend is captured by
the spherical collapse predictions (shaded band in Fig.~\ref{plot:diff.mf}).
The upper limit corresponds to unmodified forces, whereas the lower limit
corresponds to enhanced forces during the entirety of the collapse. 
The enhancement of the linear $\sigma(M)$ in $f(R)$ effectively
makes objects of the same mass less rare and causes the increase in the 
Sheth-Tormen predictions for  the exponentially 
suppressed high-mass end of the mass function ($\nu\equiv\delta_c/\sigma > 1$).
Compared to this effect, that of modifying spherical collapse parameters is much smaller.
It mainly arises from the increase in virial mass with respect to
$M_{300}$ making the same $M_{300}$ correspond to rarer virialized objects.
 In this large field limit, all but the most massive halos are better described
by the modified collapse parameters.    Moreover, for the purposes of establishing upper 
limits on $|f_{R0}|$ using the halo mass function, use of this prediction 
would only err on the conservative side.  

When the value of the $f_R$ field becomes comparable to the cosmological
potential wells, the chameleon effect starts to operate. This can be
seen in the mass function deviations for $|f_{R0}|=10^{-5}$ 
and $10^{-6}$ (see  Fig.~\ref{plot:diff.mf}). For the smallest field value,
the departures
from $\Lambda$CDM become very small, so that individual high-mass halos
change only slightly in mass. Due to the limited statistics in our simulation
sample, we are not able to reliably estimate the uncertainties on
 the mass function deviation 
for the highest mass bin in this case. However, the
mean deviation in this mass bin is consistent with zero.

The no-chameleon simulations show a behavior of increasing
deviations at high masses similar to the large-field case, while the 
full $f(R)$ simulations deviate significantly from this trend, especially
at high masses.  
For $|f_{R0}|=10^{-6}$ the excess almost entirely disappears at the 
highest masses leaving a pile-up of halos at intermediate masses.  
As in the 
power spectrum \cite{Pkpaper}, the chameleon mechanism qualitatively
changes the predictions for the mass function for $|f_{R0}| \simlt 10^{-5}$. 

\begin{figure}[tb]
   \begin{center}
     \centerline{\epsfxsize=3.4in\epsffile{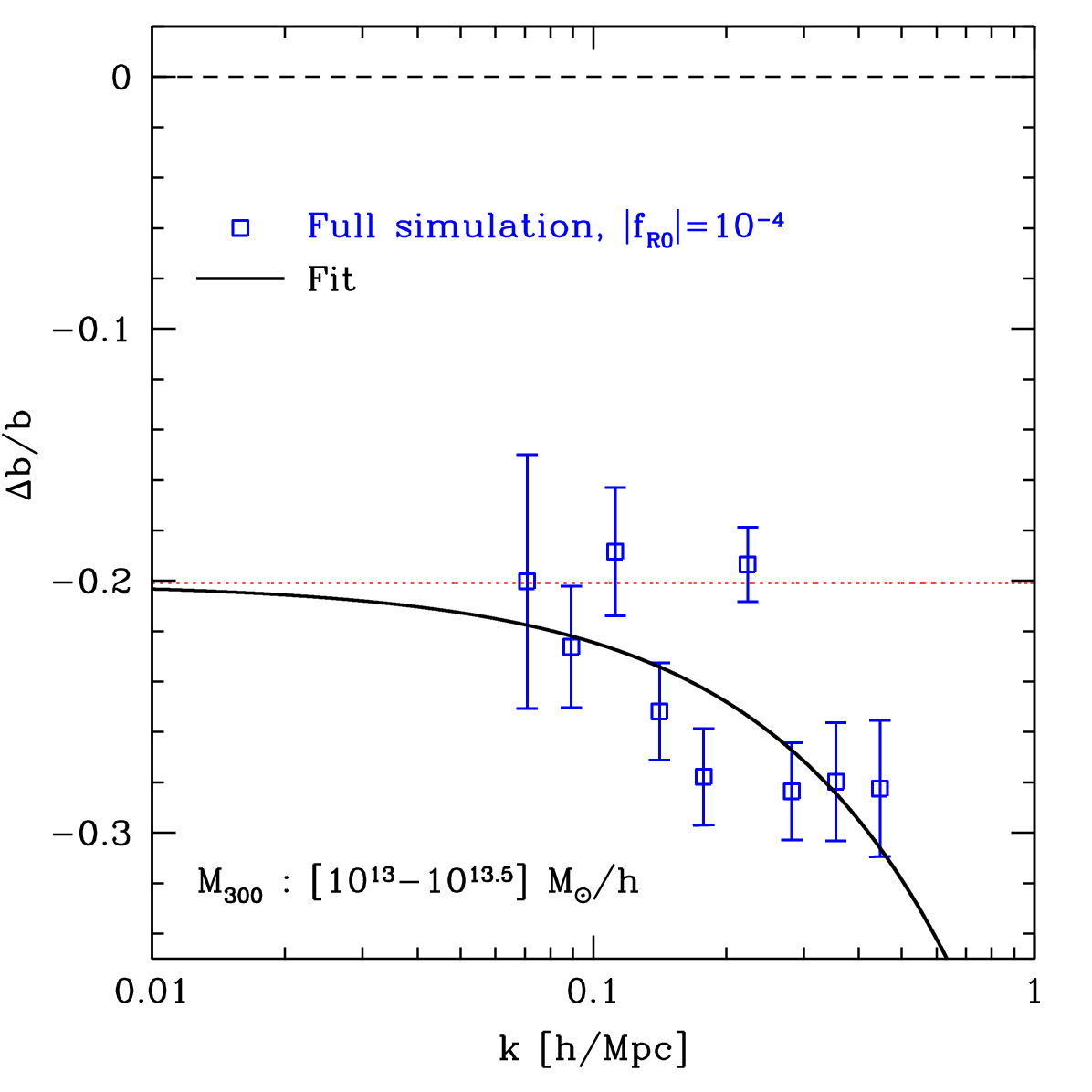}}
   \end{center}
  \caption{\footnotesize
Relative deviations in the halo bias, $\Delta b/b \equiv (b_{f(R)}-b_{\Lambda\rm CDM})/b_{\Lambda\rm CDM}$, as a function of wavenumber $k$ between $|f_{R0}|=10^{-4}$ and  $\Lambda$CDM for  $M_{300}=10^{13}-10^{13.5} h^{-1}M_{\odot}$. 
The black solid line indicates a linear fit to the bootstrap means and errors of the
combined boxes, whose extrapolation to $k=0$ 
gives $\Delta b_{\rm L}/b_{\rm L}$ (dotted red line).
} 
  \label{plot:diff.bk}
\end{figure}

\begin{figure}[htb!]
   \begin{center}
     \centerline{\epsfxsize=3.4in\epsffile{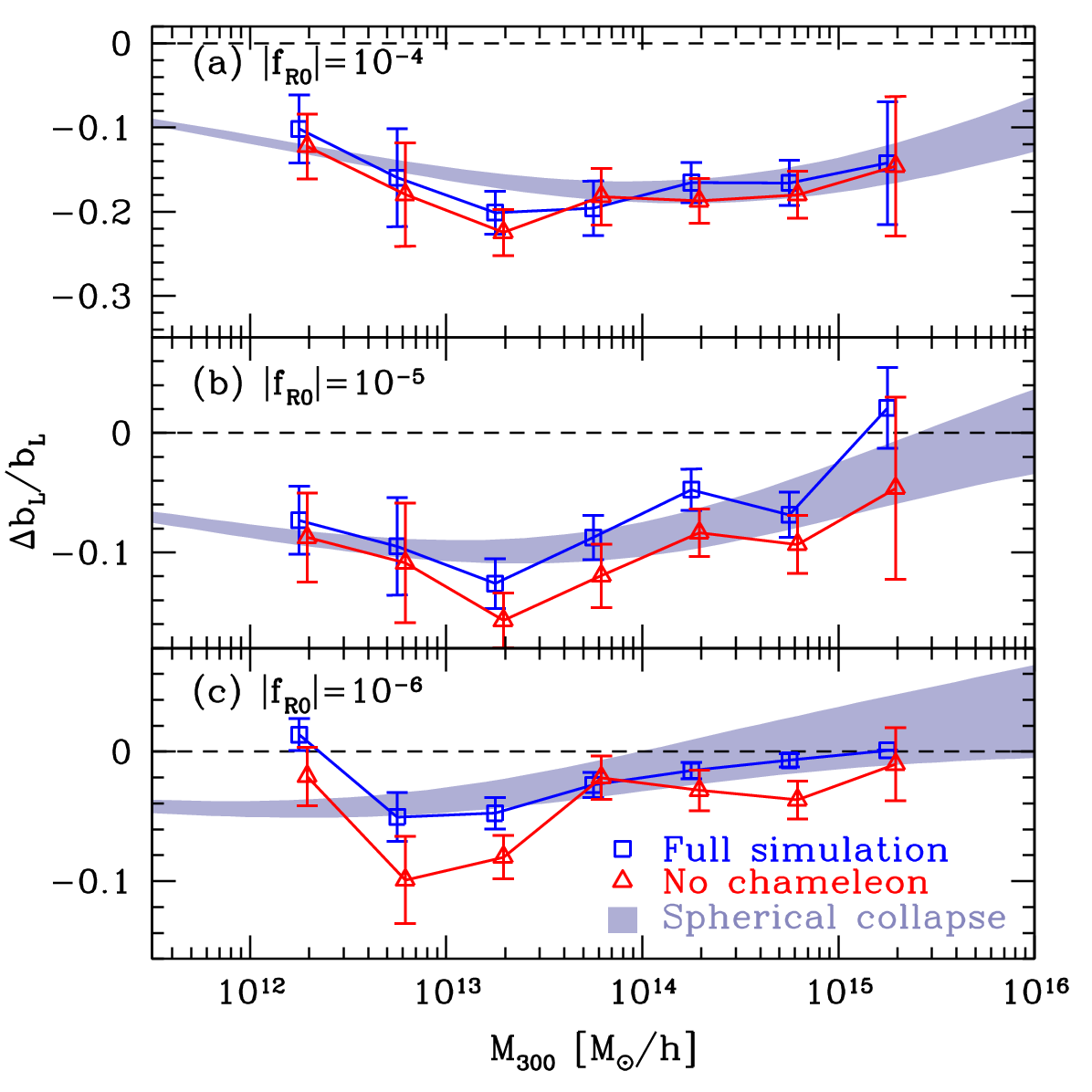}}
   \end{center}
  \caption{ \footnotesize
Relative deviations in the $f(R)$ {\it linear} halo bias from $\Lambda$CDM, 
with $|f_{R0}|=10^{-4}$ (top panel), 
$10^{-5}$ (middle panel) 
and $10^{-6}$ (lower panel).
The no chameleon simulations are again displaced horizontally for better 
visibility.
The shaded bands show the range of deviations of halo bias in $f(R)$ expected 
from spherical collapse with the upper limit corresponding to modified spherical
collapse parameters.  
} 
  \label{plot:diff.b1}
\end{figure}

It is also apparent  from Fig.~\ref{plot:diff.mf} (lower panel)
that the spherical collapse predictions are less accurate for the small field limit. 
The range of predictions encompasses a deficit of high mass halos
that is not seen in the simulations.   Since $\sigma(M)$ is calculated from the
linear prediction at a radius that encloses the mass $M$ at the background 
density, there would be no predicted enhancement of linear  fluctuations
if this radius is larger than the Compton scale in the background.  
This is in spite of the fact that
in the no-chameleon simulations forces are still enhanced once the perturbation 
collapses to smaller scales.  Combined with the rescaling of the virial mass, this
can produce a deficit of predicted objects at a fixed overdensity.  
This problem highlights the difficulties in applying scaling relations between the
linear and non-linear regime, which were developed for scale-free $\Lambda$CDM 
type models, to modified gravity theories.

In the case of the full $f(R)$ simulations, the problem is partially compensated by the appearance of the chameleon mechanism
which also reduces the abundance of the highest mass objects by eliminating the
extra force during the collapse. While the full simulation results lie within the range of spherical collapse
predictions at the high mass end, spherical collapse fails to predict the pile up of halos
at intermediate masses.

Still, the ST mass function predictions can be used to conservatively place
upper limits on $|f_{R0}|$ from the abundance of halos with $M > 10^{14}M_\odot/h$. 
Employing the modified collapse prescription for the enhancement or zero,
whichever is greater, will always underestimate the true enhancement in the
suite of models we have tested.  This underestimate becomes a small fraction of
the total enhancement for $|f_{R0}|>10^{-5}$.

\subsection{Halo Bias} \label{subsec:bias}

The halo bias computed from Eq.~(\ref{eqn:defbias}) 
in the $\Lambda$CDM simulations is shown in Fig.~\ref{plot:bkGR} for halos 
with masses in the range 
$M_{300}=10^{13}-10^{13.5}$ $h^{-1}M_{\odot}$ as an example. The points and error bars are bootstrap 
averages and errors of individual bias computations from the various boxes and 
runs. In this case, only boxes with size $L_{\rm box}=64$ and $128$ $h^{-1}$Mpc
have halos in the mass range and contribute to the bias calculation 
(see Tab.~\ref{table:runs}). Note that due to the limited halo statistics
and our small simulation sample, the scatter in the errors themselves is 
significant. We have verified that consistent results are also obtained with a
lowered $N_{\rm min} = 100-400$ which increases halo statistics allowing the larger, more linear,
boxes to be used for the bias. In the lower panel
of Fig.~\ref{plot:bkGR} we show the variation of the bias measurements with
box size. In the regime of mutual applicability, the bias measurements between boxes
are consistent within the statistical uncertainties. 

 In Fig.~\ref{plot:b1GR}, we show the linear halo bias 
in our $\Lambda$CDM simulations as a function of halo mass, measured as described
in \S \ref{subsec:halo}.  
We compare these results to the ST bias prediction of Eq.~(\ref{eqn:bias}).
We again remap the virial mass $M_{\rm v}$ to $M_{300}$
and plot the prediction for $b_{\rm L}(M_{300})$.
The simulation results are consistent within $\sim 20\%$ of
the prediction.  

\begin{figure}[tb]
  \begin{minipage}[t]{3.4in}
     \centerline{\epsfxsize=3.4in\epsffile{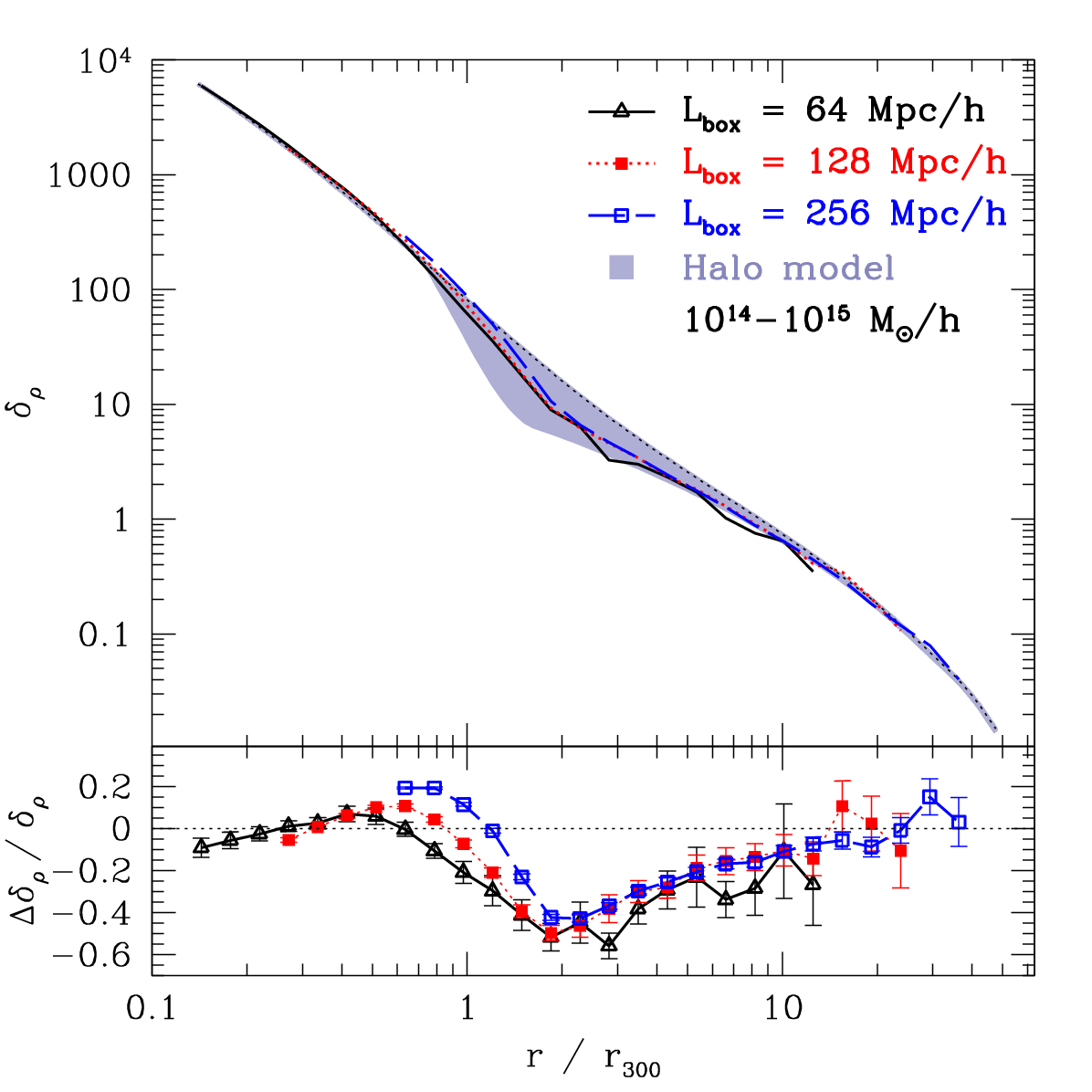}}
  \end{minipage}
  \caption{\footnotesize{
Halo density profile, expressed as the fractional overdensity
$\delta_{\rho}$, for
$M=10^{14}-10^{15}\:\Msun/h$ measured in the $\Lambda$CDM simulations
(upper panel). The halo-mass correlation predictions (shaded) represent the range with to without (dotted) profile truncation
[Eq.~(\ref{eq:xihm})] averaged over the same mass bin.
The lower panel shows the relative deviation and bootstrap errors
measured in the different boxes from 
the prediction without truncation.} 
}
  \label{plot:haloprofLCDM}
\end{figure}

\begin{figure}[tb]
  \begin{minipage}[t]{3.4in}
     \centerline{\epsfxsize=3.4in\epsffile{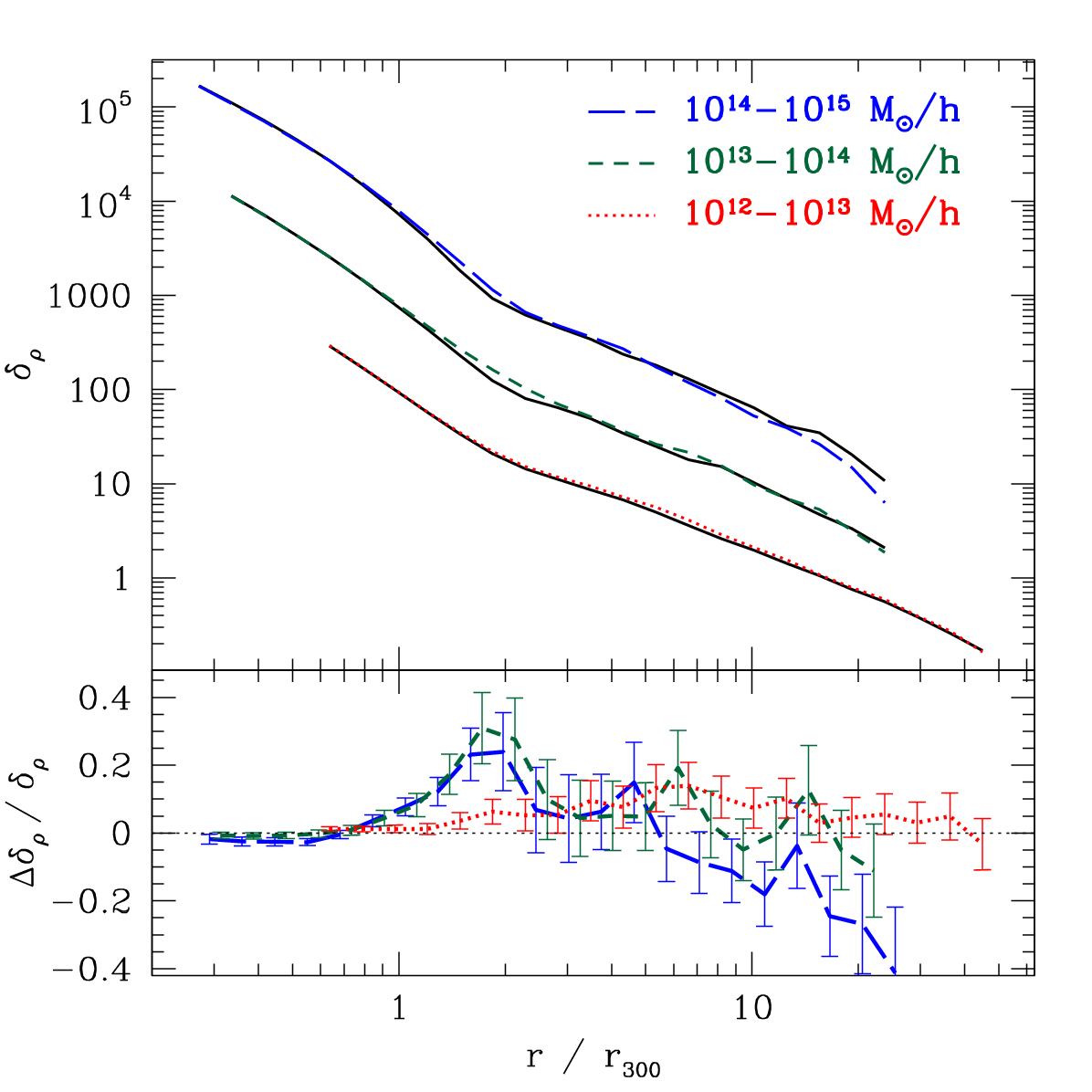}}
  \end{minipage}
  \caption{\footnotesize{
Halo density profile $\delta_\rho$ in the full $f(R)$ ($|f_{R0}|=10^{-4}$, colored) and $\Lambda$CDM simulations 
(black), for
different halo masses (upper panel). Profiles for $10^{13}-10^{14}$ and
$10^{14}-10^{15}\:\Msun/h$ have been multiplied by 10 and 100, respectively.
The profiles of the highest mass halos were obtained from $128\:\Mpch$ boxes,
while the lower mass profiles are from $64\:\Mpch$ boxes.
The lower panel shows the relative deviation of the $f(R)$ profiles from those
of $\Lambda$CDM, with bootstrap error bars.}}
  \label{plot:haloprof}
\end{figure}

Whereas the abundance of halos can be significantly changed in $f(R)$, 
their clustering properties are relatively 
less affected compared with $\Lambda$CDM.  
In Fig.~\ref{plot:diff.bk} we show the relative difference 
between the halo bias in $f(R)$ simulations with $|f_{R0}|=10^{-4}$ 
and $\Lambda$CDM for the same mass bin of Fig.~\ref{plot:bkGR}. 
For each box and run
contribution, we subtracted the $f(R)$ simulation bias from that of the 
corresponding $\Lambda$CDM simulation with same initial conditions to 
form $\Delta b(k,M)/b(k,M)$. 
The averages and error displayed are again obtained by bootstrap of the 
individual differences. 
The same linear fit procedure is applied and evaluated at $k=0$
to estimate the relative difference in the linear bias 
$\Delta b_{\rm L}(M)/b_{\rm L}(M) \equiv \Delta b/b(k,M)|_{k=0}$.
 
In Fig.~\ref{plot:diff.b1} we compare the linear bias from 
$f(R)$ and $\Lambda$CDM simulations, computed as above, and the range 
of predictions from spherical collapse.  The bias decreases with increasing 
$|f_{R0}|$ since halos of a fixed mass become less rare and thus less highly biased.  
The chameleon effect in the full simulations decreases the difference in bias versus
the no chameleon simulations
as expected.  As with the mass function, the spherical collapse range adequately describes
the high mass halos even for the small-field chameleon cases due to a fortuitous 
cancellation of modeling errors.

\subsection{Halo Profiles}
\label{subsec:prof}

The final ingredient in a basic understanding of halos and cosmological statistics
that are built out of them is their average profiles.
We plot the fractional density contrast $\delta_\rho(r/r_{300})$ defined in Eq.~(\ref{eqn:haloprofile}) and
measured in the $\Lambda$CDM simulations for the largest and hence best 
resolved mass bin in 
\refFig{haloprofLCDM} (upper panel), for different box sizes of the $\Lambda$CDM
simulations.  For reference
we compare these with the corresponding halo model prediction (shaded) from the halo-mass 
correlation function of \refeq{xihm}, consisting
of an NFW profile plus a 2-halo term describing the surrounding mass, 
averaged over the same mass bin as the simulations.  
The range of predictions shown is bounded from above by a continued 
NFW profile, and bounded from below by an NFW profile truncated at 
$r_{\rm v}=r_{390}$
as used in the halo model description of power spectra, \refeq{Pkhalo}.
In the lower panel of \refFig{haloprofLCDM}, we show the same profiles
relative to the halo model prediction with continued profiles.  Removing the
overall trend with the halo model better reveals the internal consistency of our
simulations.
The agreement between the smallest box and the larger boxes with
coarser resolution and smaller particle
number  is $\lesssim 20$\% in case of the 128$\:\Mpch$
boxes, and $\lesssim 40$\% for the 256$\:\Mpch$ boxes. In the following,
we show results from the 128$\:\Mpch$ boxes for the largest mass halos,
in order to increase halo statistics, and from the 64$\:\Mpch$ boxes for 
all other masses.

\refFig{haloprof}, top panel, shows the stacked halo profiles for three
mass bins, for $\Lambda$CDM and full $f(R)$ simulations with $|f_{R0}|=10^{-4}$. 
The lower panel of \reffig{haloprof} shows the relative deviation between
$\Lambda$CDM and $f(R)$ halo profiles.
When scaled to
the same overdensity radius, halos in $\Lambda$CDM and $f(R)$ apparently have very
similar profiles, especially in the inner part of the halo. Although a 
precise measurement of the NFW scale radius is not possible with our
limited resolution, it is apparent that there are no dramatic effects
of modified gravity on the halo concentration $c_{300}\equiv r_{300}/r_s$.
Moreover the deviations are consistent with zero well within $r_{300}$.
The same holds for the no-chameleon $f(R)$ simulations.

 For the intermediate and larger halo masses, 
there is an enhancement of the halo profile at $r/r_{300} \sim$ few, i.e.
in the transition region between one-halo and two-halo contributions.
The smallness of the enhancement of $\xi_{\rm hm}$ can be explained by
a partial cancellation between the increased linear power spectrum and reduced
linear bias in $f(R)$ (\S~\ref{subsec:scaling} and \S~\ref{subsec:bias}).
However, a quantitative understanding of the behavior of the halo-mass
correlation at these radii is not possible with the simple halo model adopted 
here, as it fails in the transition region between one and two-halo terms
(see \reffig{haloprofLCDM}).
In the small field simulations, the deviations in
the halo profiles are too small to be measured
with our current suite of simulations.


Given the relative smallness of the modified gravity effects on halo profiles,  the main effect
of enhanced forces in the large field simulations is to change the mass and hence the
abundance and bias of halos.  
\begin{figure}[tb]
  \begin{minipage}[t]{3.4in}
     \centerline{\epsfxsize=3.4in\epsffile{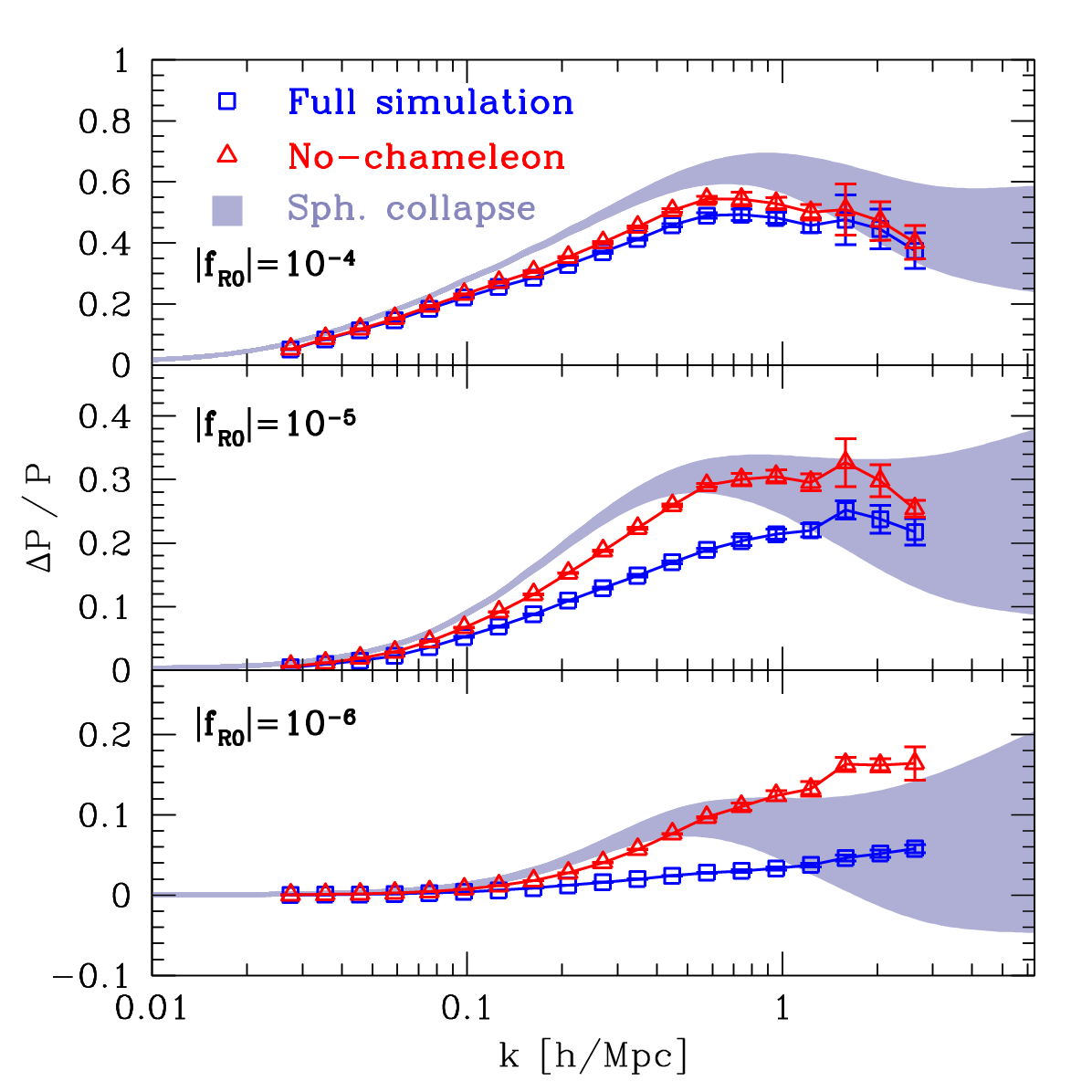}}
  \end{minipage}
  \caption{\footnotesize
Power spectrum enhancement relative to $\Lambda$CDM for 
full and no-chameleon simulation and different $f_{R0}$ field strengths. 
The shaded band shows the
predictions from the halo model using parameters derived 
from spherical collapse (see text). 
} 
  \label{plot:Pk}
\end{figure}

\subsection{Halo Model Power Spectrum}
\label{subsec:power}

We can now put the halo properties together and discuss statistics  that can be
interpreted under the  halo model paradigm outlined in \S \ref{subsec:scaling}.
The matter power spectrum $P_{\rm m m}$ is especially interesting in that 
the enhancement in the large field $f(R)$ simulations
found in  \cite{Pkpaper} was not well described by standard linear to nonlinear
scaling relations  \cite{smith03a}.  Without an adequate description of the large
field limit, robust upper limits on $|f_R|$, which should be available from current 
observations, are difficult to obtain.

The halo model provides a somewhat more physically motivated scaling relation
between the linear and nonlinear power spectra \cite{HuSaw07b}.  Specifically we use the 
same range of ST predictions for the mass function and linear bias discussed
in the previous sections in Eq.~(\ref{eq:Pkhalo}). 
In addition, we vary the concentration parameter of the halos, using either
an unmodified $c_{\rm v}(M_{\rm v})$ relation [\refeq{cvir}], or an
unmodified $c_{300}(M_{300}) \equiv r_{300}/r_{s}$. The latter relation is motivated
by our finding that the inner parts of halo profiles are unmodified in
$f(R)$ when referred to the same overdensity radius (\S~\ref{subsec:prof}).
Converting $c_{300}$ to the virial concentration, we obtain a 
$\sim$10\% higher $c_{\rm v}$, which increases the power spectrum 
enhancement at $k \gtrsim 1\:\iMpch$ through the 1-halo term [\refeq{Pkhalo}].

The range of halo model predictions is shown in \reffig{Pk} 
for different values of $f_{R0}$, 
together with the simulation results from \cite{Pkpaper}.
The upper boundary of each shaded band corresponds to unmodified spherical
collapse parameters and unchanged $c_{300}$,
while the lower boundary is using the modified spherical collapse parameters,
assuming enhanced forces throughout in the $f(R)$ prediction, and unchanged
$c_{\rm v}$. 

The halo model provides a reasonable approximation to the
relative deviations in the large field limit out to the $k \sim 1-3\:h$/Mpc scales
that can be resolved by the simulations.  The modified collapse provide
a somewhat better and more conservative approximation for the purposes
of establishing upper limits for $|f_{R0}| \simgt 10^{-4}$.

The halo model still fails to capture the chameleon suppression in the small
field limit.   Its failure is apparent even at $|f_{R0}|=10^{-5}$ for
$0.1 \simlt k (h/$Mpc) $\simlt 1$ and is relatively larger than the error in the mass 
function, linear bias and halo profiles themselves.     This range also corresponds to the regime where
the one halo and two halo terms are comparable, i.e. where our simple prescription 
of linear clustering of halos with density profiles truncated at the virial radius
cannot be expected to apply.  

A prescription that seeks to interpolate between
modified and unmodified force law predictions \cite{HuSaw07b} and a better treatment of
the transition regime that includes nonlinear halo clustering and halo exclusion
could potentially provide a better description but is beyond the scope of this study.

\section{Discussion} \label{sec:discussion}

Dark matter halos are the building blocks of cosmological observables associated
with structure in the universe.
Their statistical properties provide many interesting tests of cosmic acceleration,
especially of those that seek to modify gravitational forces.   

Here we have examined the abundance, clustering and profiles of dark matter halos
in $f(R)$ modified gravity models.   In these models, gravitational forces are enhanced
below the {\it local} Compton scale of an extra scalar degree of freedom $f_R$. 
Generically, this extra force leads to an enhanced abundance of massive halos and
a decrease in the bias of such halos, but relatively little change to the density profile
or mass correlation around halos of fixed mass.
The extent of these effects on halo statistics depends strongly on whether the background
scalar field is in the large field ($|f_{R0}|\gtrsim 10^{-5}$) or 
small field limit ($|f_{R0}|\lesssim 10^{-5}$).  

In the \textit{large field limit}, forces are modified everywhere below the {\it background} Compton 
scale ($\lambda_C\simgt 10\;\Mpch$ today \cite{Pkpaper}).  The modifications in
this regime are relatively well described by scaling relations for halo statistics.  
By modifying spherical collapse parameters to include the enhanced forces, we have
shown that the mass function and linear halo bias can be described well by the
Sheth-Tormen prescription.   The halo-mass correlation and average density profiles
are little changed from $\Lambda$CDM due to a cancellation of effects from the
enhanced forces and decreased bias.   

Together these provide a description of
the enhanced matter power spectrum that corresponds to a relatively small 
overestimate of $|f_{R0}|$ by $\sim 50\%$ or less.  This level of accuracy 
more than suffices for an order of magnitude
constraint on field values. Moreover, the overestimate depends only weakly on $|f_{R0}|$ and
can largely be corrected.  
In this prescription, concentration uncertainties which
are unresolved in our simulations should be marginalized.  
Concentration uncertainties also arise from 
 baryonic effects in $\Lambda$CDM \cite{RuddEtal} and marginalization 
over these leaves only the
 more unique intermediate scale deviations to distinguish modifications of
 gravity \cite{ZentnerEtal}.  
 
In the \textit{small field limit}, potential wells of dark matter halos are 
comparable to or larger than the background $f_R$ field, so that
the local Compton wavelength decreases substantially from the background value. 
Modifications to gravitational forces then decrease in the interior of halos by the
so-called chameleon mechanism.   
This decrease has the effect of bringing deviations in all of the halo statistics down at
the high mass end.    At intermediate masses, the excess in the halo abundance 
can actually increase further due to a pile up of halos which also suppresses the
change in the bias. 

Scaling relations are not  as easily modified to include the
chameleon effect but do still have limited applicability.  Due to a fortuitous
cancellation of problems associated with a small background Compton wavelength
and the chameleon mechanism, the modified Sheth-Tormen mass function can still
be used to provide upper limits on the field values that err only on the conservative side.
Likewise the bias description is reasonably accurate for intermediate to high mass halos. 
We caution that this fortuitous cancellation does not apply to all quantities that
can be built out of halo statistics.  For example the halo model for the power spectrum
overpredicts the enhancement in the weakly non-linear regime.    

To summarize, in the large field limit which encompasses the range that 
current cosmological observations can test, the scaling relations presented
here should already enable strong tests of the model. However, 
more work in calibrating the effects of $f(R)$ gravity will be required when 
cosmological observations reach the $\sim$10\% percent level precision 
required to test the small field limit of $f(R)$ modified gravity.

\bigskip 

\noindent {\it Acknowledgments}:  We thank N. Dalal, B. Jain, A. Kravtsov, 
U. Seljak, A. Upadhye,
and A. Vikhlinin
for useful conversations.
This work was supported in part by the Kavli Institute for Cosmological
 Physics (KICP) at the University of Chicago through grants NSF PHY-0114422 and
 NSF PHY-0551142 and an endowment from the Kavli Foundation and its 
founder Fred Kavli.
WH and ML  were additionally supported by
U.S.~Dept.\ of Energy contract DE-FG02-90ER-40560 and WH by 
 the David and Lucile Packard Foundation. 
Computational resources were provided by the KICP and by the KICP-Fermilab
computer cluster.

\appendix

\section{Spherical Collapse}
\label{app:sph_collapse}

In this Appendix, we examine the modifications to spherical collapse induced by 
the enhanced forces of the $f(R)$ model and in particular, derive the collapse threshold
$\delta_c$ and the virial overdensity $\Delta_{\rm v}$ used in the main text.

We begin with the nonlinear continuity and Euler equation for a pressureless fluid of non-relativistic
matter.   When expressed in terms of the gravitational potential $\Psi$, these equations
are unaltered by the modification to gravity that remains a metric theory (e.g.~\cite{Peebles80})
\begin{eqnarray}
{\partial \delta  \over \partial t} + {1\over a} \nabla \cdot (1+\delta) {\bf v} &=& 0\,, \nonumber\\
{\partial {\bf v} \over \partial t}  + {1\over a} ({\bf v} \cdot \nabla) {\bf v} + H {\bf v} &=& -{1\over a}\nabla \Psi\,,
\label{eqn:mattereom}
\end{eqnarray}
where $\delta = \delta \rhom/\bar \rho_{\rm m}$ and spatial coordinates are comoving.  These can be combined to a second order 
equation for $\delta$
\begin{eqnarray}
{\partial^2 \delta \over \partial t^2} + 2 H {\partial \delta \over \partial t} - {1 \over a^2} {\partial^2(1+\delta) v^{i} v^{j}  \over  \partial x^i \partial x^j} = {\nabla \cdot (1+\delta)\nabla \Psi \over a^2}
\end{eqnarray}
but require further information about the velocity and potential fields to form a closed system.

The potential is given by the field equation~(\ref{eqn:frorig}) and modified Poisson equation (\ref{eqn:potorig}) in terms of the
density fluctuation.
For the velocity field, we will take an initial top hat density perturbation and make the approximation that it remains a top hat throughout the evolution.  This approximation is valid in the limiting
cases that the Compton radius is either much larger or much smaller than the perturbation.

Given the top hat assumption for the density, the velocity field in the interior
takes the form  ${\bf v} = A(t){\bf r}$ to have a spatially constant divergence.  Its amplitude
is 
related to the top hat density perturbation through the continuity equation
(\ref{eqn:mattereom})
\begin{equation}
\dot \delta + {3\over a} (1+\delta) A =0  \,.
\end{equation}
With the relation
\begin{equation}
{\partial^2 v^iv^j\over\partial x^i \partial x^j} = 12 A^{2}  ={4\over 3}a^{2}{\dot \delta^{2} \over (1+\delta)^{2}}\,,
\end{equation}
the spherical collapse equation in the top hat approximation becomes
\begin{eqnarray}
{\partial^2 \delta \over \partial t^2} + 2 H {\partial \delta \over \partial t} - {4 \over 3} {\dot \delta^{2} \over (1+\delta)}
 &=& {(1+\delta) \over a^2}\nabla^2 \Psi  \,,
 \label{eqn:tophateqn}
\end{eqnarray}
which along with Eqs. (\ref{eqn:frorig}) and (\ref{eqn:potorig}) (\S~\ref{subsec:fr})  complete the system.

We can bring this equation to its more usual form for the radius of the top hat by
using mass conservation 
\begin{equation}
M = (4\pi /3)r^{3} \bar \rho_{\rm m} (1+\delta) = {\rm const.}
\end{equation}
 Therefore the evolution of $r$ and $\delta$ may be related as
\begin{eqnarray}
{\ddot r \over r} = {H^2 + \dot H - {1 \over 3(1+\delta)} (\ddot \delta + 2 \dot \delta H - {4 \over 3}{\dot\delta^2 \over 1+\delta }})\,.
\end{eqnarray}
Combining this relation with the top hat density equation (\ref{eqn:tophateqn}), we obtain
\begin{eqnarray}
{\ddot r \over r} = -{4\pi G \over 3} [\bar \rho_{\rm m} + (1+3w) \bar\rho_{\rm eff}] -{1\over 3a^2} \nabla^2 \Psi \,,
\end{eqnarray}
where we have expressed the background expansion in terms of an effective dark 
energy contribution.   
Note that this set of equations also applies to any smooth dark energy contribution as long as
we take $\delta R = 8\pi G \delta \rhom$ in the Poisson equation.

For the $f(R)$ system, there are two limiting cases worth noting and these both fall into
the class of top hat preserving evolution.  In the large field  case the Compton
wavelength is so long that $f_R$ ignores the collapse.  In this case $\delta R \ll 8\pi G \delta\rhom$
in the interior.   
In the opposite small field case, the Compton wavelength in the
background is always smaller than the scale of the perturbation.   In this case
$ \delta R = 8\pi G\delta \rhom$ as in ordinary gravity with smooth dark energy.
 The two limits for
the top hat equation (\ref{eqn:tophateqn}) can be parameterized
as
\begin{eqnarray}
{\ddot r \over r} = -{4\pi G \over 3} [\rho_{\rm m} + (1+3w) \bar\rho_{\rm eff}] -{4\pi G \over 3}
{F} \delta \rhom 
\end{eqnarray}
with $F=1/3$ corresponding to the large field limit and $F=0$ corresponding to the small field limit or smooth dark energy. Note that $\rhom$ in the first
term on the right hand side
stands for the total matter overdensity, so that for $F=0$ the top-hat overdensity
follows the same equation of motion as the background expansion in 
a smooth dark energy model.

We now specialize this equation for a background expansion that is close to $\Lambda$CDM,
$w=-1$ and $\bar \rho_{\rm eff}=\rho_\Lambda$. 
Rewriting the time derivatives in term of $'=d/d\ln a$, a $\Lambda$CDM background and with $y  = [r-r_i a/a_{i}]/r_i$
\begin{eqnarray}
y'' + {H' \over H} y' &=& -{1 \over 2} {\Omega_{m}a^{-3}-2 \Omega_{\Lambda} \over 
\Omega_{m} a^{-3} + \Omega_{\Lambda}}y \\
&& -{1\over 2} {\Omega_{m}a^{-3}\over
\Omega_{m} a^{-3} + \Omega_{\Lambda}}(1+{F})({a \over a_{i}}+y)\delta \nonumber
\end{eqnarray}
with
\begin{equation} 
\delta = \left({1 \over y a_i/a +1}\right)^3 (1+\delta_i) -1
\end{equation}
and $\delta_i$ as the initial density perturbation at $a_i$.  Turn around occurs when
$r'=0$ or $y' = -a/a_i$ and collapse occurs when $r=0$ or $y=-a/a_i$.  

\begin{figure}[tb]
  \begin{minipage}[t]{3.4in}
     \centerline{\epsfxsize=3.4in\epsffile{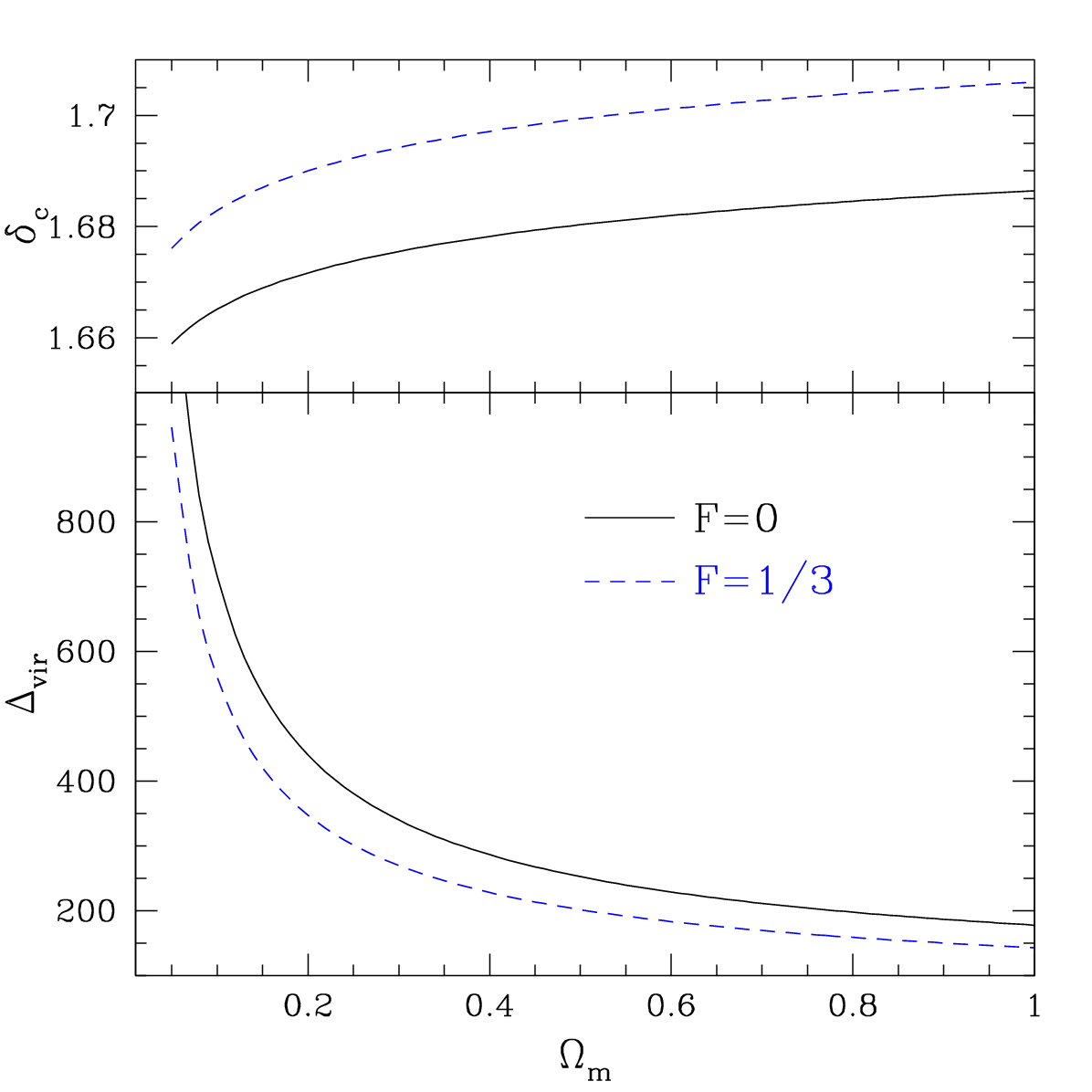}}
  \end{minipage}
  \caption{\footnotesize
Spherical collapse parameters.   The linear overdensity extrapolated to the collapse
epoch $\delta_{c}$ and the virial overdensity $\Delta_{\rm v}$ are modified from the flat
$\Lambda$CDM values ($F=0$) by the enhanced forces during collapse ($F=1/3$).
} 
  \label{plot:spherical}
\end{figure}

Under the assumption that the initial conditions are
set during matter domination  when $\delta \ll 1$, linear theory says that 
$\delta \propto a^{1+p}$ where
\begin{equation}
p = -{5\over 4} + {5 \over 4} \sqrt{ 1 + {24\over 25} F}\,.
\end{equation}
The initial conditions are then $y=0$ and $y'= -\delta_{i}(1+p)/3$.  More generally, the 
linearization of the continuity and Euler equations imply
\begin{equation}
\delta'' + 3{H' \over H}\delta' = {4\pi G \rho_m \over H^2} F\delta \,.
\end{equation}
The linear overdensity extrapolated to the collapse epoch is then a function of 
$F$.  For collapse during matter domination 
$\delta_c = 1.686$ for $F=0$ as usual and  $\delta_{c}=1.706$ for $F=1/3$.
In Fig.~\ref{plot:spherical} (lower panel), we show the threshold for collapse at $z=0$
as a function of $\Omega_{\rm m}$.  
In particular for $\Omega_{\rm m} = 0.24$,
$\delta_c = 1.673$ for $F=0$ and $\delta_c=1.692$ for $F=1/3$.

To relate spherical collapse with virialized halos, one also has to modify the virial theorem
for $f(R)$.
All the steps in the usual derivation of the tensor virial theorem from the Boltzmann equation 
still apply to $f(R)$ since the Boltzmann equation (energy momentum conservation in the metric)
is unchanged (see e.g. \cite{BinTre87}).   The only change is in relating the potential energy to the matter
in the top hat
\begin{equation}
W = -{3 \over 5} (1+F){G M^2 \over r} \,.
\end{equation}

The implications for spherical collapse then remain largely unchanged when expressed in terms
of the turn around radius.  During matter domination 
the scalar virial theorem still reads $W = -2 T$ and 
$W(r_{\rm max})= W(r_{\rm v})+T(r_{\rm v}) = {W(r_{\rm v})/2}$ and so 
$r_{\rm v} = r_{\rm max}/2$.  The difference is in
the density evolution in spherical collapse.   The traditional way of expressing the virial
overdensity $\Delta_{\rm v}$ is
to take the overdensity at $r_{\rm v}$ during the collapse $\rhom(r_{\rm v})$ and divide by the average density
at the end of collapse $\bar \rho_{\rm m}(r=0)$.   
For collapse in the matter dominated limit $F=0$ gives the usual  $\Delta_{\rm v}=177.6$ and $F=1/3$ gives
$\Delta_{\rm v}=143.1$.

These conditions are modified by the acceleration of the expansion at low redshifts.
Following \cite{Lahetal91},  the metric
effect of $\Lambda$ can be considered as providing a potential energy per unit mass of $w_\Lambda =  -4\pi G \bar \rho_{\rm eff} r^2 /3$.  Integrating this up through the top hat we get $W_\Lambda = -(4\pi G \bar \rho_{\rm eff}/5 )M r^2$.
The virial theorem with the combined potential energy gives
\begin{equation}
T = -{1\over 2} W + W_\Lambda\,.
\end{equation}
The different dependence on $r$ changes the virialization radius to the extent that $W_\Lambda$
is important.  Let us define the ratio at turnaround
\begin{equation}
\eta = {2 \rho_{\rm eff} \over (1+F) \rhom} = {2 \Omega_\Lambda \over (1+F) \Omega_m a^{-3}(1+\delta)} \,.
\end{equation}
The relationship between the virial radius and the turnaround radius $s=r_{\rm v}/r_{\rm max}$ can then be obtained from inverting
\begin{equation}
\eta = {2 s -1 \over 2s^3 - s}\,.
\end{equation}
Note that as $\eta\rightarrow 0$, $s \rightarrow 1/2$ as expected.   The effect of $F$ is to make the $\Lambda$
term less important.

In Fig.~\ref{plot:spherical} (lower panel), we show the virial overdensity for collapse at $z=0$
as a function of $\Omega_{\rm m}$.  
In particular,  for $F=0$ the virial overdensity is 
 $\Delta_{\rm v} = 390$ for collapse
today and for $F=1/3$ it is lowered to $\Delta_{\rm v}=309$.

These modifications also imply that the virial temperature of halos of a fixed
virial mass is proportional to $(1+F) \Delta_{\rm v}^{1/3}$ and hence increases for
$F=1/3$.   Likewise, hydrostatic equilibrium masses or any masses defined dynamically
by the velocity dispersion of the matter would be larger than lensing
masses by a factor of $(1+F)$.

\section{Scaling Relations}
\label{subsec:scaling} 

In this appendix, we present the scaling relations that were used for
comparisons with the simulations in section~\ref{sec:results}.
For the mass function we use
the Sheth-Tormen (ST) prescription \cite{SheTor99}.  
Though other, potentially more accurate, descriptions for
$\Lambda$CDM exist (e.g.~\cite{Tinker:2008ff}), this choice enables us 
 to explore the changes expected in the $f(R)$ simulations from
spherical collapse (see Appendix \ref{app:sph_collapse}). We also found a good match to the 
ST mass function in our $\Lambda$CDM simulations 
(\S~\ref{subsec:dndm}).

The ST description 
 for the comoving number density of halos per logarithmic interval in the virial mass $M_{\rm v}$ is given by
\begin{align}
n_{\ln M_{\rm v}} \equiv
\frac{d n}{d\ln M_{\rm v}} &= {\bar \rho_{\rm m} \over M_{\rm v}} f(\nu) {d\nu \over d\ln M_{\rm v}}\,, 
         \label{eqn:massfn}
\end{align}
where the peak threshold $\nu = \delta_c/\sigma(M_{\rm v})$ and 
\begin{eqnarray}
\nu f(\nu) = A\sqrt{{2 \over \pi} a\nu^2 } [1+(a\nu^2)^{-p}] \exp[-a\nu^2/2]\,.
\end{eqnarray}
Here
$\sigma(M)$ is the variance of the linear density field convolved with a top hat of radius $r$
that encloses $M=4\pi r^3 \bar \rho_{\rm m}/3$ at the background density
\begin{eqnarray}
\sigma^2(r) = \int \frac{d^3k}{(2\pi)^3} |\tilde{W}(kr)|^2 P_{\rm L}(k)\,,
\label{eq:sigmaR}
\end{eqnarray}
where $P_{\rm L}(k)$ is the linear power spectrum and $\tilde W$ is the Fourier transform
of the top hat window.  The normalization constant $A$ is chosen 
such that $\int d\nu f(\nu)=1$. The parameter values of $p=0.3$, $a=0.75$, and
$\delta_c=1.673$ for the spherical collapse threshold have previously been shown to 
match simulations of $\Lambda$CDM at the $10-20\%$ level. 
The virial mass is defined as the mass enclosed at 
the virial radius $r_{\rm v}$, where $\Delta_{\rm v}=390$ in the $\Lambda$CDM model.
We discuss modifications to these parameters for the $f(R)$ model in \S \ref{sec:results}.

The peak-background split for halos predicts that the linear bias of halos should be
consistent with the mass function.   For the ST mass function, the bias is given by \cite{SheTor99}
\begin{eqnarray}
b_{\rm L}(M_{\rm v}) & \equiv & b(k=0,M_{\rm v}) \nonumber\\
&=&  1 + {a \nu^2 -1 \over \delta_c}
         + { 2 p \over \delta_c [ 1 + (a \nu^2)^p]}\,.
\label{eqn:bias}
\end{eqnarray}

For the halo profiles, we take an NFW form \cite{NavFreWhi97},
\begin{equation}
\rho_{\rm NFW}(r) = \frac{\rho_s}{r/r_s (1+ r/r_s)^2},
\label{eq:NFW}
\end{equation}
where $r_s$ is the scale radius of the halo
and the normalization $\rho_s$ is given
by the virial mass $M_{\rm v}$. We parametrize $r_s$
via the concentration $c_{\rm v}\equiv r_{\rm v}/r_s$ given by 
\cite{Buletal01}:
\begin{equation}
c_{\rm v}(M_{\rm v},z=0) = 9 \left (\frac{M_{\rm v}}{M_*} \right )^{-0.13},
\label{eq:cvir}
\end{equation}
where $M_*$ is defined via $\sigma(M_*)=\delta_c$. 
By assuming an NFW form, we can also rescale mass definitions from the virial mass
$M_{\rm v}$ to $M_{300}$
as outlined in \cite{HuKravtsov}.  We use this approach 
to compare these scaling relation predictions to the simulations in \S \ref{sec:results}
since the definition of the virial mass varies with cosmological parameters and $f(R)$ modifications.
For a given halo in $\Lambda$CDM, $M_{300}$ is slightly larger than $M_{\rm v}$.
Given that we generally rescale to $M_{300}$,
when no specific overdensity is given we implicitly take $M=M_{300}$,
e.g. 
\begin{equation}
n_{\ln M} \equiv {d n \over d{\ln M_{300}}} = n_{\ln M_{\rm v}} {d{\ln M_{\rm v}}  \over d{\ln M_{300}}} \,.
\end{equation}

These properties are combined together in the halo model which treats cosmological
statistics associated with structures through the halos that form them (see \cite{CooraySheth}
for a review).
For example, the matter power spectrum can be
decomposed into  1-halo and 2-halo terms,
\begin{eqnarray}
P_{\rm mm}(k) &=&I^2(k) P_{\rm L}(k)  + P^{1{\rm h}}(k)\,,\nonumber\\
P^{1{\rm h}}(k) &=& \int d\ln M_{\rm v}\: n_{\ln M_{\rm v}} \frac{M_{\rm v}^2}{\bar{\rho}_{\rm m}^2} |y(k,M_{\rm v})|^2\,,
\label{eq:Pkhalo}
\end{eqnarray}
where
\begin{equation}
I(k) = 
 \int d\ln M_{\rm v}\: n_{\ln M_{\rm v}} \frac{M_{\rm v}}{\bar{\rho}_{\rm m}} y(k,M_{\rm v}) b_{\rm L}(M_{\rm v}) \,.
\end{equation}
Here, $y(k,M)$ is the Fourier transform of an NFW density profile truncated at $r_{\rm v}$, unless otherwise specified,
and normalized
so that $y(k,M)\rightarrow 1$ as $k\rightarrow 0$.   Note that with the ST mass function and 
bias, $\lim_{k \rightarrow 0} I(k) = 1$.

Likewise the halo-mass cross spectrum $P_{\rm hm}$ for an infinitesimally narrow mass bin around $M_{\rm v}$ is given by
\begin{eqnarray}
P_{\rm hm} &=& b_L(M_{\rm v}) I(k) P_{\rm L}(k) + {M_{\rm v} \over \bar \rho_{\rm m}} y(k,M_{\rm v}) \,.
\label{eq:Phm}
\end{eqnarray}
Note that the Fourier transform of this quantity is the halo-mass correlation 
function, or average mass profile
\begin{eqnarray}
\xi_{\rm hm}(r) &\equiv&\frac{\langle \rho_{\rm h}(r)\rangle }{\bar\rho_{\rm m}} -1 = \int {d^3 k \over (2\pi)^3} 
P_{\rm hm} e^{-i {\bf k}\cdot {\bf x}} \nonumber\,, \\
&=& b_{\rm L}(M_{\rm v}) \int {d^3 k \over (2\pi)^3}  I(k) P_{\rm L}(k) e^{-i {\bf k}\cdot {\bf x}} \nonumber\\
&& \quad +  {\rho_{\rm NFW}(r)\over \bar \rho_{\rm m}} \,.
\label{eq:xihm}
\end{eqnarray}
For comparison with simulations, we show the $\rho_{\rm NFW}$ term with and without
the truncation at the virial radius in \S \ref{subsec:prof}.  Both the overly simplistic
treatment of halo profiles and the use of linear halo correlations make our simple model
inaccurate in the region where the one and two halo pieces are comparable.

\bibliography{halopaper}

\begin{thebibliography}{28}
\expandafter\ifx\csname natexlab\endcsname\relax\def\natexlab#1{#1}\fi
\expandafter\ifx\csname bibnamefont\endcsname\relax
  \def\bibnamefont#1{#1}\fi
\expandafter\ifx\csname bibfnamefont\endcsname\relax
  \def\bibfnamefont#1{#1}\fi
\expandafter\ifx\csname citenamefont\endcsname\relax
  \def\citenamefont#1{#1}\fi
\expandafter\ifx\csname url\endcsname\relax
  \def\url#1{\texttt{#1}}\fi
\expandafter\ifx\csname urlprefix\endcsname\relax\def\urlprefix{URL }\fi
\providecommand{\bibinfo}[2]{#2}
\providecommand{\eprint}[2][]{\url{#2}}

\bibitem[{\citenamefont{Sotiriou and Faraoni}(2008)}]{Sotiriou:2008rp}
\bibinfo{author}{\bibfnamefont{T.~P.} \bibnamefont{Sotiriou}} \bibnamefont{and}
  \bibinfo{author}{\bibfnamefont{V.}~\bibnamefont{Faraoni}}
  (\bibinfo{year}{2008}), \eprint{0805.1726}.

\bibitem[{\citenamefont{Nojiri and Odintsov}(2008)}]{Nojiri:2008nt}
\bibinfo{author}{\bibfnamefont{S.}~\bibnamefont{Nojiri}} \bibnamefont{and}
  \bibinfo{author}{\bibfnamefont{S.~D.} \bibnamefont{Odintsov}}
  (\bibinfo{year}{2008}), \eprint{0807.0685}.

\bibitem[{\citenamefont{Carroll et~al.}(2004)\citenamefont{Carroll, Duvvuri,
  Trodden, and Turner}}]{Caretal03}
\bibinfo{author}{\bibfnamefont{S.~M.} \bibnamefont{Carroll}},
  \bibinfo{author}{\bibfnamefont{V.}~\bibnamefont{Duvvuri}},
  \bibinfo{author}{\bibfnamefont{M.}~\bibnamefont{Trodden}}, \bibnamefont{and}
  \bibinfo{author}{\bibfnamefont{M.~S.} \bibnamefont{Turner}},
  \bibinfo{journal}{Phys. Rev.} \textbf{\bibinfo{volume}{D70}},
  \bibinfo{pages}{043528} (\bibinfo{year}{2004}), \eprint{astro-ph/0306438}.

\bibitem[{\citenamefont{Nojiri and Odintsov}(2003)}]{NojOdi03}
\bibinfo{author}{\bibfnamefont{S.}~\bibnamefont{Nojiri}} \bibnamefont{and}
  \bibinfo{author}{\bibfnamefont{S.~D.} \bibnamefont{Odintsov}},
  \bibinfo{journal}{Phys. Rev.} \textbf{\bibinfo{volume}{D68}},
  \bibinfo{pages}{123512} (\bibinfo{year}{2003}), \eprint{hep-th/0307288}.

\bibitem[{\citenamefont{Capozziello et~al.}(2003)\citenamefont{Capozziello,
  Carloni, and Troisi}}]{Capozziello:2003tk}
\bibinfo{author}{\bibfnamefont{S.}~\bibnamefont{Capozziello}},
  \bibinfo{author}{\bibfnamefont{S.}~\bibnamefont{Carloni}}, \bibnamefont{and}
  \bibinfo{author}{\bibfnamefont{A.}~\bibnamefont{Troisi}},
  \bibinfo{journal}{Recent Res. Dev. Astron. Astrophys.}
  \textbf{\bibinfo{volume}{1}}, \bibinfo{pages}{625} (\bibinfo{year}{2003}),
  \eprint{astro-ph/0303041}.

\bibitem[{\citenamefont{{Khoury} and {Weltman}}(2004)}]{khoury04a}
\bibinfo{author}{\bibfnamefont{J.}~\bibnamefont{{Khoury}}} \bibnamefont{and}
  \bibinfo{author}{\bibfnamefont{A.}~\bibnamefont{{Weltman}}},
  \bibinfo{journal}{\prd} \textbf{\bibinfo{volume}{69}},
  \bibinfo{pages}{044026} (\bibinfo{year}{2004}),
  \eprint{arXiv:astro-ph/0309411}.

\bibitem[{\citenamefont{Cembranos}(2006)}]{Cembranos:2005fi}
\bibinfo{author}{\bibfnamefont{J.~A.~R.} \bibnamefont{Cembranos}},
  \bibinfo{journal}{Phys. Rev.} \textbf{\bibinfo{volume}{D73}},
  \bibinfo{pages}{064029} (\bibinfo{year}{2006}), \eprint{gr-qc/0507039}.

\bibitem[{\citenamefont{Navarro and Van~Acoleyen}(2007)}]{Navarro:2006mw}
\bibinfo{author}{\bibfnamefont{I.}~\bibnamefont{Navarro}} \bibnamefont{and}
  \bibinfo{author}{\bibfnamefont{K.}~\bibnamefont{Van~Acoleyen}},
  \bibinfo{journal}{JCAP} \textbf{\bibinfo{volume}{0702}}, \bibinfo{pages}{022}
  (\bibinfo{year}{2007}), \eprint{gr-qc/0611127}.

\bibitem[{\citenamefont{Faulkner et~al.}(2007)\citenamefont{Faulkner, Tegmark,
  Bunn, and Mao}}]{Faulkner:2006ub}
\bibinfo{author}{\bibfnamefont{T.}~\bibnamefont{Faulkner}},
  \bibinfo{author}{\bibfnamefont{M.}~\bibnamefont{Tegmark}},
  \bibinfo{author}{\bibfnamefont{E.~F.} \bibnamefont{Bunn}}, \bibnamefont{and}
  \bibinfo{author}{\bibfnamefont{Y.}~\bibnamefont{Mao}},
  \bibinfo{journal}{Phys. Rev.} \textbf{\bibinfo{volume}{D76}},
  \bibinfo{pages}{063505} (\bibinfo{year}{2007}), \eprint{astro-ph/0612569}.

\bibitem[{\citenamefont{{Hu} and {Sawicki}}(2007{\natexlab{a}})}]{HuSaw07a}
\bibinfo{author}{\bibfnamefont{W.}~\bibnamefont{{Hu}}} \bibnamefont{and}
  \bibinfo{author}{\bibfnamefont{I.}~\bibnamefont{{Sawicki}}},
  \bibinfo{journal}{\prd} \textbf{\bibinfo{volume}{76}},
  \bibinfo{pages}{064004} (\bibinfo{year}{2007}{\natexlab{a}}),
  \eprint{arXiv:0705.1158}.

\bibitem[{\citenamefont{Oyaizu}(2008)}]{oyaizu08b}
\bibinfo{author}{\bibfnamefont{H.}~\bibnamefont{Oyaizu}},
  \bibinfo{journal}{\prd} \textbf{\bibinfo{volume}{\rm submitted}}
  (\bibinfo{year}{2008}), \eprint{0807.2449}.

\bibitem[{\citenamefont{{Oyaizu} et~al.}(2008)\citenamefont{{Oyaizu}, {Lima},
  and {Hu}}}]{Pkpaper}
\bibinfo{author}{\bibfnamefont{H.}~\bibnamefont{{Oyaizu}}},
  \bibinfo{author}{\bibfnamefont{M.}~\bibnamefont{{Lima}}}, \bibnamefont{and}
  \bibinfo{author}{\bibfnamefont{W.}~\bibnamefont{{Hu}}},
  \bibinfo{journal}{\prd} \textbf{\bibinfo{volume}{\rm submitted}}
  (\bibinfo{year}{2008}), \eprint{0807.2462}.

\bibitem[{\citenamefont{{Brandt}}(1973)}]{brandt73}
\bibinfo{author}{\bibfnamefont{A.}~\bibnamefont{{Brandt}}},
  \bibinfo{journal}{{Proceedings of Third International Conference on Numerical
  Methods in Fluid Mechanics}}  (\bibinfo{year}{1973}).

\bibitem[{\citenamefont{Briggs et~al.}(2000)\citenamefont{Briggs, Henson, and
  McCormick}}]{briggs00a}
\bibinfo{author}{\bibfnamefont{W.~L.} \bibnamefont{Briggs}},
  \bibinfo{author}{\bibfnamefont{V.~E.} \bibnamefont{Henson}},
  \bibnamefont{and} \bibinfo{author}{\bibfnamefont{S.~F.}
  \bibnamefont{McCormick}}, \emph{\bibinfo{title}{A multigrid tutorial (2nd
  ed.)}} (\bibinfo{publisher}{Society for Industrial and Applied Mathematics},
  \bibinfo{address}{Philadelphia, PA, USA}, \bibinfo{year}{2000}), ISBN
  \bibinfo{isbn}{0-89871-462-1}.

\bibitem[{\citenamefont{{Jenkins} et~al.}(2001)\citenamefont{{Jenkins},
  {Frenk}, {White}, {Colberg}, {Cole}, {Evrard}, {Couchman}, and
  {Yoshida}}}]{Jenkins01}
\bibinfo{author}{\bibfnamefont{A.}~\bibnamefont{{Jenkins}}},
  \bibinfo{author}{\bibfnamefont{C.~S.} \bibnamefont{{Frenk}}},
  \bibinfo{author}{\bibfnamefont{S.~D.~M.} \bibnamefont{{White}}},
  \bibinfo{author}{\bibfnamefont{J.~M.} \bibnamefont{{Colberg}}},
  \bibinfo{author}{\bibfnamefont{S.}~\bibnamefont{{Cole}}},
  \bibinfo{author}{\bibfnamefont{A.~E.} \bibnamefont{{Evrard}}},
  \bibinfo{author}{\bibfnamefont{H.~M.~P.} \bibnamefont{{Couchman}}},
  \bibnamefont{and}
  \bibinfo{author}{\bibfnamefont{N.}~\bibnamefont{{Yoshida}}},
  \bibinfo{journal}{\mnras} \textbf{\bibinfo{volume}{321}},
  \bibinfo{pages}{372} (\bibinfo{year}{2001}), \eprint{arXiv:astro-ph/0005260}.

\bibitem[{\citenamefont{{Sheth} and {Tormen}}(1999)}]{SheTor99}
\bibinfo{author}{\bibfnamefont{R.}~\bibnamefont{{Sheth}}} \bibnamefont{and}
  \bibinfo{author}{\bibfnamefont{B.}~\bibnamefont{{Tormen}}},
  \bibinfo{journal}{\mnras} \textbf{\bibinfo{volume}{308}},
  \bibinfo{pages}{119} (\bibinfo{year}{1999}).

\bibitem[{\citenamefont{Tinker et~al.}(2008)}]{Tinker:2008ff}
\bibinfo{author}{\bibfnamefont{J.~L.} \bibnamefont{Tinker}}
  \bibnamefont{et~al.} (\bibinfo{year}{2008}), \eprint{0803.2706}.

\bibitem[{\citenamefont{Navarro et~al.}(1997)\citenamefont{Navarro, Frenk, and
  White}}]{NavFreWhi97}
\bibinfo{author}{\bibfnamefont{J.~F.} \bibnamefont{Navarro}},
  \bibinfo{author}{\bibfnamefont{C.~S.} \bibnamefont{Frenk}}, \bibnamefont{and}
  \bibinfo{author}{\bibfnamefont{S.~D.~M.} \bibnamefont{White}},
  \bibinfo{journal}{Astrophys. J.} \textbf{\bibinfo{volume}{490}},
  \bibinfo{pages}{493} (\bibinfo{year}{1997}), \eprint{astro-ph/9611107}.

\bibitem[{\citenamefont{{Bullock} et~al.}(2001)\citenamefont{{Bullock},
  {Kolatt}, {Sigad}, {Somerville}, {Kravtsov}, {Klypin}, {Primack}, and
  {Dekel}}}]{Buletal01}
\bibinfo{author}{\bibfnamefont{J.~S.} \bibnamefont{{Bullock}}},
  \bibinfo{author}{\bibfnamefont{T.~S.} \bibnamefont{{Kolatt}}},
  \bibinfo{author}{\bibfnamefont{Y.}~\bibnamefont{{Sigad}}},
  \bibinfo{author}{\bibfnamefont{R.~S.} \bibnamefont{{Somerville}}},
  \bibinfo{author}{\bibfnamefont{A.~V.} \bibnamefont{{Kravtsov}}},
  \bibinfo{author}{\bibfnamefont{A.~A.} \bibnamefont{{Klypin}}},
  \bibinfo{author}{\bibfnamefont{J.~R.} \bibnamefont{{Primack}}},
  \bibnamefont{and} \bibinfo{author}{\bibfnamefont{A.}~\bibnamefont{{Dekel}}},
  \bibinfo{journal}{\mnras} \textbf{\bibinfo{volume}{321}},
  \bibinfo{pages}{559} (\bibinfo{year}{2001}), \eprint{arXiv:astro-ph/9908159}.

\bibitem[{\citenamefont{{Hu} and {Kravtsov}}(2003)}]{HuKravtsov}
\bibinfo{author}{\bibfnamefont{W.}~\bibnamefont{{Hu}}} \bibnamefont{and}
  \bibinfo{author}{\bibfnamefont{A.~V.} \bibnamefont{{Kravtsov}}},
  \bibinfo{journal}{\apj} \textbf{\bibinfo{volume}{584}}, \bibinfo{pages}{702}
  (\bibinfo{year}{2003}), \eprint{arXiv:astro-ph/0203169}.

\bibitem[{\citenamefont{{Cooray} and {Sheth}}(2002)}]{CooraySheth}
\bibinfo{author}{\bibfnamefont{A.}~\bibnamefont{{Cooray}}} \bibnamefont{and}
  \bibinfo{author}{\bibfnamefont{R.}~\bibnamefont{{Sheth}}},
  \bibinfo{journal}{\physrep} \textbf{\bibinfo{volume}{372}},
  \bibinfo{pages}{1} (\bibinfo{year}{2002}), \eprint{arXiv:astro-ph/0206508}.

\bibitem[{\citenamefont{{Smith} et~al.}(2003)\citenamefont{{Smith}, {Peacock},
  {Jenkins}, {White}, {Frenk}, {Pearce}, {Thomas}, {Efstathiou}, and
  {Couchman}}}]{smith03a}
\bibinfo{author}{\bibfnamefont{R.~E.} \bibnamefont{{Smith}}},
  \bibinfo{author}{\bibfnamefont{J.~A.} \bibnamefont{{Peacock}}},
  \bibinfo{author}{\bibfnamefont{A.}~\bibnamefont{{Jenkins}}},
  \bibinfo{author}{\bibfnamefont{S.~D.~M.} \bibnamefont{{White}}},
  \bibinfo{author}{\bibfnamefont{C.~S.} \bibnamefont{{Frenk}}},
  \bibinfo{author}{\bibfnamefont{F.~R.} \bibnamefont{{Pearce}}},
  \bibinfo{author}{\bibfnamefont{P.~A.} \bibnamefont{{Thomas}}},
  \bibinfo{author}{\bibfnamefont{G.}~\bibnamefont{{Efstathiou}}},
  \bibnamefont{and} \bibinfo{author}{\bibfnamefont{H.~M.~P.}
  \bibnamefont{{Couchman}}}, \bibinfo{journal}{\mnras}
  \textbf{\bibinfo{volume}{341}}, \bibinfo{pages}{1311} (\bibinfo{year}{2003}),
  \eprint{arXiv:astro-ph/0207664}.

\bibitem[{\citenamefont{{Hu} and {Sawicki}}(2007{\natexlab{b}})}]{HuSaw07b}
\bibinfo{author}{\bibfnamefont{W.}~\bibnamefont{{Hu}}} \bibnamefont{and}
  \bibinfo{author}{\bibfnamefont{I.}~\bibnamefont{{Sawicki}}},
  \bibinfo{journal}{\prd} \textbf{\bibinfo{volume}{76}},
  \bibinfo{pages}{104043} (\bibinfo{year}{2007}{\natexlab{b}}),
  \eprint{arXiv:0708.1190}.

\bibitem[{\citenamefont{{Rudd} et~al.}(2008)\citenamefont{{Rudd}, {Zentner},
  and {Kravtsov}}}]{RuddEtal}
\bibinfo{author}{\bibfnamefont{D.~H.} \bibnamefont{{Rudd}}},
  \bibinfo{author}{\bibfnamefont{A.~R.} \bibnamefont{{Zentner}}},
  \bibnamefont{and} \bibinfo{author}{\bibfnamefont{A.~V.}
  \bibnamefont{{Kravtsov}}}, \bibinfo{journal}{\apj}
  \textbf{\bibinfo{volume}{672}}, \bibinfo{pages}{19} (\bibinfo{year}{2008}),
  \eprint{arXiv:astro-ph/0703741}.

\bibitem[{\citenamefont{{Zentner} et~al.}(2008)\citenamefont{{Zentner}, {Rudd},
  and {Hu}}}]{ZentnerEtal}
\bibinfo{author}{\bibfnamefont{A.~R.} \bibnamefont{{Zentner}}},
  \bibinfo{author}{\bibfnamefont{D.~H.} \bibnamefont{{Rudd}}},
  \bibnamefont{and} \bibinfo{author}{\bibfnamefont{W.}~\bibnamefont{{Hu}}},
  \bibinfo{journal}{\prd} \textbf{\bibinfo{volume}{77}},
  \bibinfo{pages}{043507} (\bibinfo{year}{2008}), \eprint{0709.4029}.

\bibitem[{\citenamefont{{Peebles}}(1980)}]{Peebles80}
\bibinfo{author}{\bibfnamefont{P.~J.~E.} \bibnamefont{{Peebles}}},
  \emph{\bibinfo{title}{{The large-scale structure of the universe}}}
  (\bibinfo{publisher}{Research supported by the National Science
  Foundation.~Princeton, N.J., Princeton University Press, 1980.~435 p.},
  \bibinfo{year}{1980}).

\bibitem[{\citenamefont{Binney and Tremaine}(1987)}]{BinTre87}
\bibinfo{author}{\bibfnamefont{J.}~\bibnamefont{Binney}} \bibnamefont{and}
  \bibinfo{author}{\bibfnamefont{S.}~\bibnamefont{Tremaine}},
  \emph{\bibinfo{title}{Galactic Dynamics}} (\bibinfo{publisher}{Princeton
  University Press}, \bibinfo{year}{1987}).

\bibitem[{\citenamefont{{Lahav} et~al.}(1991)\citenamefont{{Lahav}, {Lilje},
  {Primack}, and {Rees}}}]{Lahetal91}
\bibinfo{author}{\bibfnamefont{O.}~\bibnamefont{{Lahav}}},
  \bibinfo{author}{\bibfnamefont{P.~B.} \bibnamefont{{Lilje}}},
  \bibinfo{author}{\bibfnamefont{J.~R.} \bibnamefont{{Primack}}},
  \bibnamefont{and} \bibinfo{author}{\bibfnamefont{M.~J.}
  \bibnamefont{{Rees}}}, \bibinfo{journal}{\mnras}
  \textbf{\bibinfo{volume}{251}}, \bibinfo{pages}{128} (\bibinfo{year}{1991}).

\end{thebibliography}

\end{document}